\documentclass[aps,prmaterials,twocolumn,groupedaddress,showpacs,notitlepage]{revtex4-1} 
\usepackage{graphicx}
\usepackage{subfigure}
\usepackage{amsmath}
\usepackage{braket}
\usepackage[colorlinks,linkcolor={blue},citecolor={blue},urlcolor={blue}]{hyperref}

\usepackage{array}
\usepackage{MnSymbol}

\usepackage{bbm} 
\usepackage{chngcntr} 
\usepackage[page,toc,titletoc,title]{appendix}

\usepackage{array}
\newcolumntype{P}[1]{>{\centering\arraybackslash}p{#1}}
\newcolumntype{M}[1]{>{\centering\arraybackslash}m{#1}}

\newcommand{\bA}{\mathbf{A}}
\newcommand{\bB}{\mathbf{B}}
\newcommand{\br}{\mathbf{r}}
\newcommand{\bk}{\mathbf{k}}
\newcommand{\bp}{\mathbf{p}}
\newcommand{\bG}{\mathbf{G}}
\newcommand{\bL}{\mathbf{L}}
\newcommand{\bK}{\mathbf{K}}
\newcommand{\ba}{\mathbf{a}}
\newcommand{\bb}{\mathbf{b}}
\newcommand{\bS}{\mathbf{S}}
\newcommand{\bdelta}{\boldsymbol{\delta}}
\newcommand{\bsigma}{\boldsymbol{\sigma}}

\newcommand{\bl}{\mathbf{l}}

\begin{document}

\title{Theory of tunable flux lattices in the homobilayer moir\'e of twisted and uniformly strained transition metal dichalcogenides}

\author{Dawei Zhai}
\email[]{dzhai@hku.hk}

\author{Wang Yao}
\email[]{wangyao@hku.hk}

\affiliation{Department of Physics and HKU-UCAS Joint Institute of Theoretical and Computational Physics at Hong Kong, The University of Hong Kong, Hong Kong, China}

\date{\today}

\begin{abstract}
The spatial texture of internal degree of freedom of electrons has profound effects on the properties of materials. Such texture in real space can manifest as an emergent magnetic field (or Berry curvature), which is expected to induce interesting valley/spin-related transport phenomena. Moir\'e pattern, which emerges as a spatial variation at the interface of 2D atomic crystals, provides a natural platform for investigating such real space Berry curvature effects. Here we study moir\'e structures formed in homobilayer transition metal dichalcogenides (TMDs) due to twisting, various uniform strain profiles, and their combinations, where electrons can reside in either layer with the layer index serving as an internal degree of freedom. The layer pseudo-spin exhibits vortex/antivortex textures in the moir\'e supercell, leading to a giant geometric magnetic field and a scalar potential. Within a geometric picture, the moir\'e magnetic field is found as the cross product of the gradients of the out-of-plane pseudo-spin and the in-plane pseudo-spin orientation respectively. We discover dual roles of uniform strain: Besides being a cause of the moire atomic texture in the homobilayer, it also contributes a pseudo-gauge potential that modifies the local phase of interlayer coupling. 
Consequently, strain can be employed to tune the in-plane pseudo-spin texture, while interlayer bias tunes the out-of-plane pseudo-spin, and we show how the moir\'e magnetic field's spatial profile, intensity, and flux per supercell can be engineered.
Through the geometric scalar correction, the landscape of the scalar potential can also be engineered along with the moir\'e magnetic field, forming distinct effective lattice structures. These properties render TMD moir\'e structures promising to build tunable flux lattices for transport and topological material applications.
\end{abstract}

\maketitle

\section{Introduction}
van der Waals structures built from combining various 2D materials with different electronic and optical properties have attracted intense research interests in recent years.\cite{Geimlegos,KostyaLegos,GraphenehBNNatPhysRev2019,OscarReview,JustinSongNatNano2018} Among various heterostructure geometries, vertically stacked bilayers, where moir\'e patterns may emerge due to the inevitable lattice constant mismatch and/or interlayer misorientation, have been studied the most. 
The spatially modulated interfacial interactions in the moir\'e patterns endow these composite materials with novel properties and allow the observation of exciting physical phenomena that are absent in the monolayers. 
In homobilayers, small twisting between the layers, as well as spatially uniform strain applied differently on the two layers (also referred as the hetero-strain) are usually exploited to engineer long period moir\'e patterns.
Arguably, the most prominent example is twisted bilayer graphene at magic angles with flat bands,\cite{MacDonaldPNAS2011,LocalizationTwistedGrapheneNanoLett2010,TightBindingTwistedGraphenePRB2010,ContinuumModelTwistedGraphenePRL2007,JeilJungAbinitioTwistedGraphenePRB2014} where exotic superconducting and correlated insulating states have been observed.\cite{CaoYuanSuperconductivity,CaoYuanCorrelated,CoryDeanSuperconductivityScience,FerromagnetismTwistedBilayerGrapheneScience,MagnetismTwistedBilayerGrapheneNature} Moir\'e structures formed by transition metal dichalcogenides (TMDs) also receive significant attentions, especially towards their optical signatures (e.g. moir\'e exciton) because of their semiconductor nature.\cite{XiaodongXuMoireExcitonNature2019,XiaoqinLiMoireExcitonNature2019,FengWangMoireExcitonNature2019,FalkoMoireExcitonNature2019,HongyiMoireExcitonSciAdv2017,FengchengWuMoireExcitonPRL2017,FengchengWuMoireExcitonPRB2018}

Compared to their monolayer counterparts, bilayer moir\'e structures exhibit two extra characteristics. The first is the layer pseudo-spin internal degree of freedom (DoF) since particles can reside in either layer, and the pseudo-spin configuration can depend on the stacking order.\cite{HongyiPseudoFieldMoire,WuMacDonaldPRL2019} The second is spatial variation  
of local stacking configurations in each moir\'e unit cell.\cite{FlatBandSolitonTwistedTMDPRL2018} Their coexistence implies spatially modulated pseudo-spin internal DoF. It is well established that non-trivial spatial texture of the internal DoF has profound effects on electronic properties, which can be understood in terms of Berry curvature and Berry phase.\cite{QianNiuRMP} In real space, texture of the internal DoF is manifested as an emergent magnetic field, which might induce valley/spin Hall effects in 2D materials. 
Previously, such real space Berry phase effects have been mostly modelled by utilising optical lattice schemes.\cite{NonAbelianGaugeRMP,GoldmanNonAbelianReview,GeometricGaugeFieldOpticalLatticeNCooper} While it is clear from the above discussions that moir\'e serves as such a platform naturally without requiring complicated setups. 

In this work, we present a systematic study on the emergent magnetic field arising from the moir\'e patterns introduced by twisting and uniform strain in TMD homobilayers. 
The intralayer potential and interlayer coupling in the moir\'e together manifest as a spatially varying pseudo-spin Zeeman field $\mathcal{\vec{V}}$ (defined in Eq.~(\ref{Eq:ValenceBandMoirePotential})) that couples to the layer pseudospin-$\frac{1}{2}$ DoF,\cite{WuMacDonaldPRL2019} giving rise to vortex/antivortex textures of pseudospin orientations. We show that effects of such non-trivial spatial texture can be reformulated in terms of a non-Abelian gauge potential, which leads to a giant geometric magnetic field and a scalar potential in the adiabatic limit when the particle dynamics is projected onto either branch of the pseudo-spin eigenstates.
The magnetic field and scalar potential are geometric in nature because they depend on the spatial variations of the spherical angles of $\mathcal{\vec{V}}$,\cite{NonAbelianGaugeRMP} thus should be distinguished from similar quantities emerging in other systems, e.g. inhomogeneously strained 2D crystals without moir\'e.\cite{StrainPhysRep2016,StrainPhysRep2010,FujitaGaugeFieldReview,SasakiStrainGraphene,ZhaiStrainMPLB}
We give a geometric relation where the emergent magnetic field is expressed as the cross product of the gradients of the out-of-plane and in-plane pseudo-spin textures.
While the out-of-plane pseudo-spin is coupled to the interlayer bias, we find the spatial profile of in-plane pseudo-spin orientation responds to the uniaxial or shear hetero-strain,
as the latter effectively introduces a pseudo-gauge potential that modifies the interlayer coupling.
One can therefore engineer the in-plane and out-of-plane layer pseudo-spin texture by employing uniform strain and interlayer bias, respectively. This allows great flexibility in tuning the profile of the magnetic field. The vortex/anti-vortex pseudo-spin texture also ensures that the magnetic flux per supercell is always quantized in a general moir\'e pattern formed with twisting and various strain profiles. Topological phase transition, i.e. sign change of the magnetic flux, may occur from the twist dominated regime to the strain dominated one. 
Furthermore, landscape of the scalar potential also changes accordingly by tuning the twist angle and strain, which acts as a guide to form distinct effective lattice structures in the tight-binding limit to characterize various moir\'e profiles.\cite{WuMacDonaldPRL2019,HongyiPseudoFieldMoire,DecoratedTriangularLatticeTopologyChange} 

\section{Continuum valence band model of twisted homobilayer TMD}
\begin{figure}[ht]
	\includegraphics[width=2.5in]{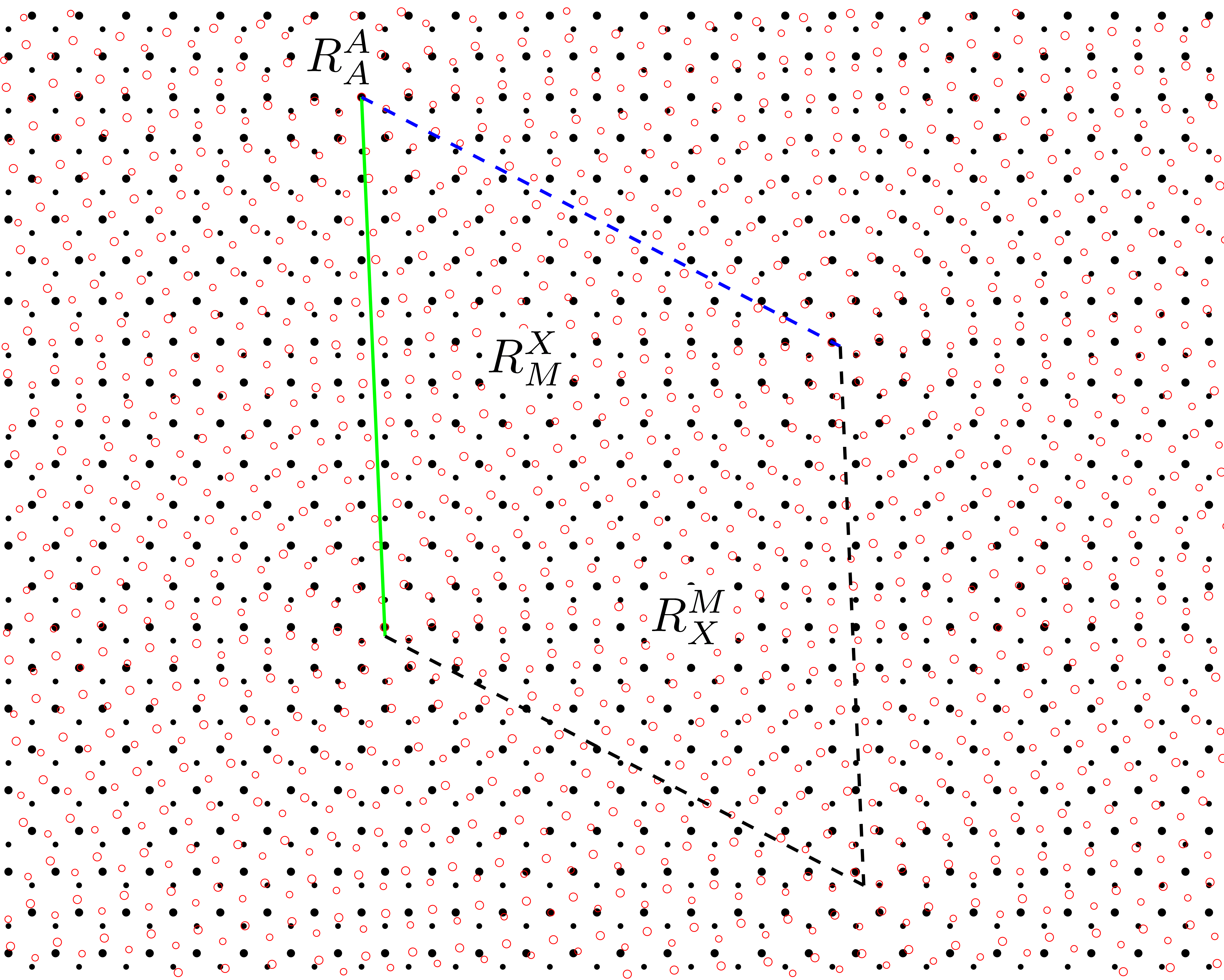}
	\caption{Moir\'e pattern formed via rotating the top layer (red) counterclockwise by $5^\circ$. The green solid line and blue dashed line represents $\bL_1$ and $\bL_2$, respectively. The parallelogram encloses one moir\'e unit cell. High symmetry local stackings are indicated as well. Larger (smaller) symbols denote $M$ ($X$) atoms.}
	\label{Fig:MoireRotation}
\end{figure}

In the following, we will focus on moir\'e built from parallelly stacked (or R-stacking) homobilayer TMD, and use parameters of the MoSe$_2$ compound.\cite{HongyiPseudoFieldMoire} 
Consider a homobilayer with the top layer rotated counterclockwise by an angle $\theta$. A moir\'e pattern will form with local high symmetry stacking configurations: $R^A_A$, $R^M_X$, and $R^X_M$ (Fig.~\ref{Fig:MoireRotation}). $R^A_A$ represents aligned parallel stacking, and $R^i_j$ represents Bernal stacking with $i$ atoms from top layer sit above $j$ atoms from the bottom layer. Here $M$ ($X$) represents metal (chalcogen) atoms, and $R$ indicates R-stacking.

The twisting is characterized by the rotation matrix $R(\theta)$.
A vector $\br_0$ in the top layer is changed into $\br=R\br_0$, and the corresponding displacement is $\bdelta(\br)=\br-\br_0=\left(\mathbbm{1}-R^{-1}\right)\br$,
where $\mathbbm{1}$ is the identity matrix.
The moir\'e primitive lattice vectors $\bL_{i=1,2}$ can be defined via $\bdelta(\bL_i)=\ba_i$, where $\ba_i$ is the primitive lattice vector of monolayer TMD that are chosen as $\ba_1=\left(1,0\right)a$ and $\ba_2=\left(1/2,\sqrt{3}/2\right)a$, and $a$ is the lattice constant. Here we assign the zigzag (armchair) edge as the $x$ $(y)$ axis. Also, we assume that a metal atom sits at the origin.
Therefore, the moir\'e primitive lattice vectors $\bL_i$ are given by $\bL_i=\left(\mathbbm{1}-R^{-1}\right)^{-1}\ba_i$. \cite{NanotubeMoirePRB2015}
The corresponding reciprocal lattice vectors of the moir\'e read $\bG_i=\left(\mathbbm{1}-R\right)\bb_i$,
where $\bb_1=(1,\,-1/\sqrt{3})2\pi/a$ and $\bb_2=(0,\,2/\sqrt{3})2\pi/a$ are the reciprocal lattice vectors of the monolayer.\cite{NanotubeMoirePRB2015,InterlayerCouplingNewJPhys2015} Note that one can rewrite $\bG_i$ as $\bG_i=\bb_i-\tilde{\bb}_i$, where $\tilde{\bb}_i=R\bb_i$ is the reciprocal lattice vector of the rotated layer.
The $K$ points of the two monolayers can be chosen respectively as
$\tilde{\bK}_{\tau}=\tau\left(2\tilde{\bb}_1+\tilde{\bb}_2\right)/3=R\bK_{\tau}$ and $\bK_{\tau}=\tau\left(2\bb_1+\bb_2\right)/3$,
where $\tau=\pm$ is the valley index. It can be shown that the relative shift of the $K$ points between the two layers is $\tilde{\bK}_{\tau}-\bK_{\tau}=-\tau(2\bG_1+\bG_2)/3$. As the two valleys are related by time-reversal symmetry, we will concentrate on the $K$ valley in the following, and the valley index will be neglected.

\begin{figure}[ht]
	\includegraphics[width=3in]{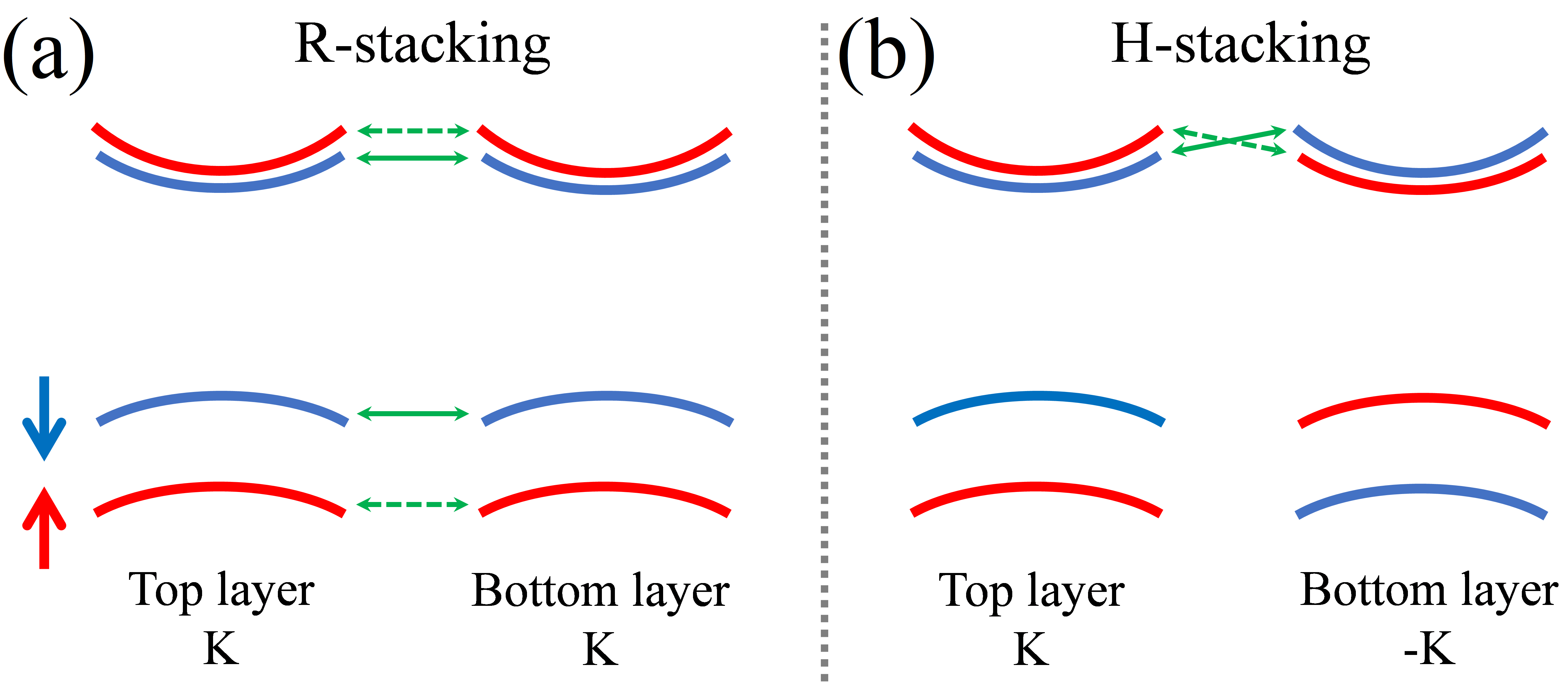}
	\caption{Schematics illustrating the different coupling scenarios in R-stacking (a) or H-stacking (b) homobilayer MoSe$_2$ with SOC-induced conduction and valence band splitting. The red and blue colors denote spin up and down, respectively. The green arrows represent interlayer coupling. We focus on the spin-valley locked valence band edges (lower blue curves in (a)) in this work.}
	\label{Fig:Rstacking_vs_Hstacking}
\end{figure}

The conduction and valence bands of TMD are separated by a large energy gap,\cite{DiXiaoTMDPRL2012,SpinSplitingBilayerTMDNatComm2013} thus the interband coupling between the two layers can be neglected. In the following, we will focus on the valence bands and consider the effects of intraband coupling between the two layers. As spin-orbit coupling (SOC) induced splitting is large in the valence bands, there exists the so-called spin-valley locking in the low energy regime, with spin up/down tied to valley $-K$/$K$.\cite{TMDChemSocRev2015} In the case of moir\'e formed from R-stacking, interlayer coupling occurs between Dirac cones with the same spin and valley indices from the two layers (Fig.~\ref{Fig:Rstacking_vs_Hstacking}(a) shows the situation at the $K$ valley). The effective Hamiltonian governing the valence bands at $K$ valley with spin down reads \cite{WuMacDonaldPRL2019} 
\begin{equation}
H_v=\left(-\frac{p^2}{2m^{*}}-\frac{E_g}{2}\right)\mathbbm{1}+\mathcal{U}_{v},\label{Eq:OriginalHamiltonian}
\end{equation}
where $m^{*}=\frac{E_g}{2v_F^2}$ is the effective mass with monolayer Fermi velocity $v_F$ and energy gap $E_g$ (Parameters are taken from Ref.~\cite{DiXiaoTMDPRL2012}).
For reference, the appendices provide details of the four-band model taking into account both conduction and valence bands (Appendix~\ref{App:FourBandContinuumModel}), and the derivation of the two-band model (Appendix~\ref{App:DeriveTwoBandModel}).
The term in the bracket describes the quadratic dispersion near the valence band edge with maximum located at $-E_g/2$. The other term
$
\mathcal{U}_{v}=
\begin{pmatrix}
V^t_v&\tilde{U}_{vv}\\
\tilde{U}_{vv}^{*}&V^b_v
\end{pmatrix}
$
is the moir\'e potential characterizing the coupling in the valence ($v$) bands with $t/b$ labeling the top/bottom layer. The corresponding eigenstate will be denoted as 
$\ket{\Psi_v}=(\Psi_{tv},\,\Psi_{bv})^T$.  The two-component form originates from the fact that particles can reside in either layer with the layer internal DoF.

The moir\'e potential $\mathcal{U}_v$ depends on the interlayer registry. For aligned bilayers, the registry is described by the constant displacement $\bdelta_0$ between the two layers, hence the potential reads $\mathcal{U}_v(\bdelta_0)$, which is uniform in space.\cite{InterlayerCouplingTMDPRB2017,WuMacDonaldPRL2019} In the case of twisted bilayers, the displacement $\bdelta(\br)$ exhibits spatial variations as shown in Fig.~\ref{Fig:MoireRotation}. If the moir\'e period is large, the local displacement varies smoothly, one can then adapt the local approximation assuming that each local pattern is approximated as an aligned configuration obtained by translation of one layer with respect to the other. One then replaces $\bdelta_0$ with $\bdelta(\br)$ in the moir\'e coupling, i.e. $\mathcal{U}_v=\mathcal{U}_v(\bdelta)$.\cite{InterlayerCouplingTMDPRB2017,WuMacDonaldPRL2019,MacDonaldPNAS2011}

Let us now discuss the intralayer moir\'e potentials $V_v^t$ and $V_v^b$.\cite{HongyiPseudoFieldMoire,InterlayerCouplingTMDPRB2017,WuMacDonaldPRL2019} They describe the band edge shifts in each individual layer due to the interlayer coupling, and they are modeled by
\begin{equation}
\begin{aligned}
V_v^t&=V_0\sum_{i=1}^{3} \cos\left(\bb_i\cdot\bdelta+\alpha\right)+V_1\\
V_v^b&=V_0\sum_{i=1}^{3} \cos\left(\bb_i\cdot\bdelta-\alpha\right)+V_1
\end{aligned},\label{Eq:IntralayerPotential}
\end{equation}
where $V_0\approx8.586$ meV, $V_1\approx-0.667$ meV, $\alpha\approx-0.49\pi$ for $\text{MoSe}_2$ obtained by fitting to first-principles results,
\footnote{Our definition of fitting parameters are related to those in Ref.~\cite{HongyiPseudoFieldMoire} via $V_1=-\frac{2\delta_0}{3}$, $V_0=\sqrt{\left(\frac{2\delta_0}{9}\right)^2+\left(\frac{2\sqrt{3}\Delta_0}{9}\right)^2}$, $\alpha=\arctan\left(-\frac{\sqrt{3}\Delta_0}{\delta_0}\right)$}
$\bb_3=-\bb_1-\bb_2$, and $\bdelta$ is the displacement vector from twisting. These expressions are consistent with those in Ref.~\cite{WuMacDonaldPRL2019} apart from the presence of $V_1$, which is negligible. 
Some of their symmetry properties worth mentioning. For a mirror reflection in the $z$ direction, the two layers are interchanged with the replacement of $\bdelta\rightarrow-\bdelta$. This implies that $V_v^t(\bdelta)=V_v^b(-\bdelta)$ and it is indeed satisfied.\cite{WuMacDonaldPRL2019}
As will be discussed in Eq.~(\ref{Eq:ValenceBandMoirePotential}), they can also be decomposed as $V_v^t=\mathcal{V}_0+\mathcal{V}_z$, and $V_v^b=\mathcal{V}_0-\mathcal{V}_z$. Since $\mathcal{V}_z$ acts as a staggered potential, it will exhibit opposite signs in regions where the local stacking configuration is flipped, e.g. $R^M_X$ $vs$ $R^X_M$.\cite{HongyiPseudoFieldMoire} This property is important to achieve (non-zero) flux lattices by employing homobilayer TMD.\cite{GeometricGaugeFieldOpticalLatticeNCooper,GeometricGaugeFieldOpticalLatticeNewJPhys,SpielmanFluxLatticeComment}

$\tilde{U}_{vv}$ describes the valence band coupling between the two layers.\cite{InterlayerCouplingTMDPRB2017,HongyiPseudoFieldMoire,WuMacDonaldPRL2019}
Here it is modeled as (see Appendix~\ref{App:FourBandContinuumModel} for details)
\begin{equation}
	\tilde{U}_{vv}=U_{vv}e^{i\frac{\theta}{2}}e^{i\left(\bK-\tilde{\bK}\right)\cdot\br},
\end{equation}
where $e^{i\frac{\theta}{2}}$ is caused by rotation of the Pauli matrices, $e^{i\left(\bK-\tilde{\bK}\right)\cdot\br}$ originates from the relative shift of the Dirac points in the two layers, and
\begin{equation}
	U_{vv}=\left(\sum_{i=1}^{3}h_0e^{i\bK_i\cdot\bdelta}+h_1e^{-i2\bK_i\cdot\bdelta}\right)e^{-i\bK_1\cdot\bdelta},
\end{equation}
with $h_0=7.1$ meV and $h_1=-1.2$ meV for $\text{MoSe}_2$.\cite{HongyiPseudoFieldMoire}
$\bK_{1-3}$ are the three equivalent Dirac points of the monolayer with $\bK_1$ selected as $\bK_1=(2\bb_1+\bb_2)/3$. \footnote{Note that Ref.~\cite{HongyiPseudoFieldMoire} defines the armchair crystalline direction as the $x$ axis, while here $x$ is along the zigzag direction. So our $\bK_1$ is related to that in Ref.~\cite{HongyiPseudoFieldMoire} by a $90^\circ$ rotation} The term associated with $h_1$ is higher order correction, which is added to better fit the DFT results.\cite{HongyiPseudoFieldMoire} Eliminating it do not affect any conclusion of the work. Actually, $U_{vv}\approx \left(\sum_{i=1}^{3}h_0e^{i\bK_i\cdot\bdelta}\right)e^{-i\bK_1\cdot\bdelta}=h_0\left[1+e^{-i\bb_1\cdot\bdelta}+e^{-i\left(\bb_1+\bb_2\right)\cdot\bdelta}\right]$ is consistent with that in twisted bilayer graphene,\cite{InterlayerCouplingNewJPhys2015,WannierTBG,NanotubeMoirePRB2015,PseudoFieldTwistedBilayerDaiXi} if one notices that $\bb_i\cdot\bdelta=\bG_{i}\cdot\br$.\cite{NanotubeMoirePRB2015,LiangFuStrainMoire,WannierTBG} 
Also note that $e^{-i\bK_1\cdot\bdelta}=e^{-i(2\bG_1+\bG_2)\cdot\br/3}=e^{i(\tilde{\bK}-\bK)\cdot\br}$, hence $\tilde{U}_{vv}$ can be rewritten more conveniently as 
\begin{equation}
\tilde{U}_{vv}=\left(\sum_{i=1}^{3}h_0e^{i\bK_i\cdot\bdelta}+h_1e^{-i2\bK_i\cdot\bdelta}\right)e^{i\frac{\theta}{2}}.\label{Eq:TildeUvv}
\end{equation}
At locations corresponding to $R^X_M$ or $R^M_X$ stacking, $\tilde{U}_{vv}$ vanishes due to the three-fold rotational symmetry.\cite{WuMacDonaldPRL2019,HongyiPseudoFieldMoire}


Note that although we only consider the valence band coupling in a moir\'e formed with parallelly stacked homobilayer MoSe$_2$ in this work, the formalism can be straightforwardly generalised to the conduction bands, as well as to other TMD compounds, including anti-parallelly stacked (or H-stacking) bilayers. The anti-parallel alignment in H-stacking indicates that the coupling should occur between different valleys of the two layers (Fig.~\ref{Fig:Rstacking_vs_Hstacking}(b)).\cite{SpinSplitingBilayerTMDNatComm2013,TMDChemSocRev2015} Since the interlayer hopping conserves spin, the large energy offset between the valence band edges of the same spin index from the two layers (about $180$ meV for MoSe$_2$) suppresses their interlayer coupling. Nevertheless, at the conduction band edges, interlayer coupling is still allowed due to the relatively small energy offset (about $20$ meV for MoSe$_2$), and the offset can also be compensated by a modest interlayer bias. With these changes taken into account, one can straightforwardly apply the formalism in our work to study H-stacking twisted bilayers.

\section{Emergence of non-Abelian gauge potential}\label{Sect:NonAbelianGauge}
The moir\'e potential $\mathcal{U}_{v}$ can be rewritten in a more physically transparent form:
\begin{equation}
\begin{aligned}
\mathcal{U}_{v}
&=
\frac{V^t_v+V^b_v}{2}\mathbbm{1}
+
\begin{pmatrix}
\frac{V^t_v-V^b_v}{2}&\tilde{U}_{vv}\\
\tilde{U}_{vv}^{*}&-\frac{V^t_v-V^b_v}{2}
\end{pmatrix}\\
&=\mathcal{V}_0\mathbbm{1}+\vec{\sigma}\cdot\vec{\mathcal{V}}
\end{aligned},\label{Eq:ValenceBandMoirePotential}
\end{equation}
where we have defined $\mathcal{V}_0=\left(V^t_v+V^b_v\right)/2$, $\mathcal{V}_x=\text{Re}\tilde{U}_{vv}^{*}$, $\mathcal{V}_y=\text{Im}\tilde{U}_{vv}^{*}$, and $\mathcal{V}_z=\left(V^t_v-V^b_v\right)/2$. Note that here $\vec{\sigma}=(\sigma_x,\,\sigma_y,\,\sigma_z)$ and $\vec{\mathcal{V}}=(\mathcal{V}_x,\,\mathcal{V}_y,\,\mathcal{V}_z)$ are three-component vectors. Overhead arrows have been used to differentiate them from two-component vectors in bold, e.g. $\bsigma$. Here $\vec{\sigma}$ represents the layer pseudo-spin, i.e. particles in a moir\'e lattice can reside in either layer, with the layer index acting as the internal DoF. $\vec{\mathcal{V}}$ behaves like an effective Zeeman field that couples to the layer pseudo-spin.
In the following, we will denote $\vec{\sigma}\cdot\vec{\mathcal{V}}$ as $\mathcal{U}_{\text{ps}}$-- the pseudo-spin coupling potential. Incidentally, many physical systems exhibit similar coupling terms,\cite{GeometricGaugeFieldOpticalLatticeNCooper,GeometricGaugeFieldOpticalLatticeNewJPhys,GoldmanNonAbelianReview,NonAbelianGaugeRMP,NonAbelianGaugeTMDCurvature} for instance, $\mathcal{U}_{\text{ps}}$ resembles the coupling between a two-level atom and laser beams, where the electronic excited and ground states denote the internal degrees of freedom (\textit{vs} top and bottom layers in the moir\'e) coupled by the laser fields (\textit{vs} moir\'e potentials here).\cite{NonAbelianGaugeRMP,GoldmanNonAbelianReview}

As $\vec{\mathcal{V}}$ exhibits spatial variation, it can be parameterized using spherical coordinates. To be specific, one can write $\vec{\mathcal{V}}=\mathcal{V}\vec{n}=\mathcal{V}\left(\sin\zeta\cos\phi^*,\,\sin\zeta\sin\phi^*,\,\cos\zeta\right)$. The amplitude $\mathcal{V}$ is defined as $\mathcal{V}=\sqrt{\left|\tilde{U}_{vv}\right|^2+\mathcal{V}_z^2}$, the polar angle $\zeta$ satisfies $\cos\zeta=\mathcal{V}_z/\mathcal{V}$ and $\sin\zeta=\left|\tilde{U}_{vv}\right|/\mathcal{V}$, and the azimuthal angle $\phi^*$ is the phase of $\tilde{U}_{vv}^{*}$, i.e. $\tilde{U}_{vv}^{*}=\left|\tilde{U}_{vv}\right|e^{i\phi^*}$. With these parameterizations, $\mathcal{U}_{\text{ps}}$ becomes
\begin{equation}
\begin{aligned}
\mathcal{U}_{\text{ps}}
=\mathcal{V}\vec{\sigma}\cdot\vec{n}
=
\mathcal{V}
\begin{pmatrix}
\cos\zeta&e^{-i\phi^*}\sin\zeta\\
e^{i\phi^*}\sin\zeta&-\cos\zeta
\end{pmatrix}
\end{aligned}.
\end{equation}
Formally, $\mathcal{U}_{\text{ps}}$ describes a pseudo-spin $1/2$ particle moving in an effective inhomogeneous Zeeman field $\vec{\mathcal{V}}$.\cite{WuMacDonaldPRL2019,GeometricGaugeFieldOpticalLatticeNCooper} Pseudo-spin can orient parallel or anti-parallel to the Zeeman field's direction $\vec{n}$, with position-dependent energy separation between them. In the following, we will explore the consequences resulting from the nontrivial spatial variations in the moir\'e lattice.\cite{GoldmanNonAbelianReview,NonAbelianGaugeRMP,HongyiPseudoFieldMoire,WuMacDonaldPRL2019} As will be shown, when a particle moves in real space, its pseudo-spin travels on the Bloch sphere defined by the spherical angles $(\zeta,\,\phi^*)$. Consequently, the particle will gain a geometric phase of $\Omega/2$, where $\Omega$ is the solid angle subtended by the trajectory of the pseudo-spin.\cite{GeometricGaugeFieldOpticalLatticeNCooper,SpielmanFluxLatticeComment} This phase can be interpreted as arising from a real space geometric gauge potential and the associated magnetic field.\cite{GeometricGaugeFieldOpticalLatticeNCooper,NonAbelianGaugeRMP,GoldmanNonAbelianReview,SpielmanFluxLatticeComment}

The pseudo-spin coupling matrix $\mathcal{U}_{\text{ps}}$ is responsible for the dynamics of the layer pseudo-spin internal DoF.
Its local eigenvectors at point $\br$ read
\footnote{At $R^X_M$ or $R^M_X$ stacking locals, $\tilde{U}_{vv}$ vanishes due to three-fold rotational symmetry, then $\phi^{*}$ is not well defined. This causes singularities in the gauge potential and is the origin of non-vanishing magnetic flux. If $\mathcal{V}_z$ happened to vanish as well (e.g. in the presence of interlayer bias), it occurs that $\epsilon_{+}=\epsilon_{-}=0$. Then Eqs.~(\ref{Eq:InternalDegreeEigenvector}) are invalid, and $\ket{\chi_{\pm}}$ can be chosen arbitrarily, e.g. $(1,\,0)^T$ and $(0,\,1)^T$, respectively.}
\begin{equation}
\begin{aligned}
\ket{\chi_+}=
\begin{pmatrix}
\cos\left(\frac{\zeta}{2}\right)e^{-i\frac{\phi^*}{2}}\\
\sin\left(\frac{\zeta}{2}\right)e^{i\frac{\phi^*}{2}}
\end{pmatrix},\,
\ket{\chi_-}=
\begin{pmatrix}
\sin\left(\frac{\zeta}{2}\right)e^{-i\frac{\phi^*}{2}}\\
-\cos\left(\frac{\zeta}{2}\right)e^{i\frac{\phi^*}{2}}
\end{pmatrix}
\end{aligned}\label{Eq:InternalDegreeEigenvector}
\end{equation}
with $\epsilon_{\pm}=\pm\mathcal{V}$ the corresponding eigenvalues. Ranges of $\zeta$ and $\phi^*$ depend on the details of the moir\'e potential. If $\zeta$ and $\phi^{*}$ span $[0,\,\pi]\times[0,\,2\pi]$, then they define the Bloch sphere (this is indeed the case for the twisted bilayer in Fig.~\ref{Fig:PseudospinAndMagneticField2deg}(b)). The north and south pole corresponds to pseudo-spin up and down, respectively.
By evaluating the pseudo-spin distribution associated with these two characteristic internal states $\braket{\vec{\sigma}}_{\pm}=\braket{\chi_{\pm}|\vec{\sigma}|\chi_{\pm}}$, one obtains 
$\braket{\vec{\sigma}}_{\pm}=\pm\vec{n}$.
Therefore, the pseudo-spin $\braket{\vec{\sigma}}_{\pm}$ has unit magnitude pointing along $\pm\vec{n}$ (Fig.~\ref{Fig:PseudospinAndMagneticField2deg}(b)), which is expected as $\mathcal{U}_{\text{ps}}\propto\vec{\sigma}\cdot\vec{n}$ projects the pseudo-spin towards $\vec{n}$. Let us stress here that $\vec{n}$ not only determines the orientation of layer pseudo-spin $\braket{\vec{\sigma}}_{\pm}$, it also maps real space position $\br=(x,\,y)$ onto the Bloch sphere surface defined by $(\zeta,\,\phi^*)$.


These two internal states are orthonormal and form a complete basis for the Hilbert space associated with the internal DoF.
Therefore, the eigenvector of the moir\'e $\ket{\Psi_v}$ can be expressed in terms of $\ket{\chi_{\pm}}$ as $\ket{\Psi_v}=\sum_{i=\pm}\tilde{\Psi}_v^i(\br)\ket{\chi_i(\br)}$,
where $\tilde{\Psi}_v^i(\br)$ is a space-dependent function characterizing the center-of-mass motion of the $i$-th internal state.\cite{GoldmanNonAbelianReview,NonAbelianGaugeRMP}

In the following, we will switch to a space-modulated spinor basis such that the $z$ axis is always along $\vec{n}$ and show that the effect of the moir\'e potential can be understood in terms of a non-Abelian gauge potential. This is achieved by applying a space-dependent unitary transformation $Q=\left(\chi_+\,\chi_-\right)$, whose columns are $\ket{\chi_{\pm}}$. It is obvious that $Q^{\dagger}\mathcal{U}_{\text{ps}}Q=\text{diag}(\epsilon_+,\,\epsilon_-)=\mathcal{V}\sigma_{z}$. $Q$ also connects $\ket{\chi_{\pm}}$ with the layer pseudo-spin basis $\ket{+}=(1,\,0)^T$ and $\ket{-}=(0,\,1)^T$, i.e. $\ket{\chi_+}=Q\ket{+}$ and $\ket{\chi_-}=Q\ket{-}$. Apply $Q^{\dagger}$ to $H_v\ket{\Psi_v}=E\ket{\Psi_v}$, we arrive at $H_v^{\text{eff}}\ket{\tilde{\Psi}_v}=E\ket{\tilde{\Psi}_v}$, where the new Hamiltonian reads
\begin{equation}
H_v^{\text{eff}}=Q^{\dagger}H_vQ=-\frac{\left(\bp+\bA\right)^2}{2m^{*}}+\left(-\frac{E_g}{2}+\mathcal{V}_0+\mathcal{V}\sigma_{z}\right),
\end{equation}
and $\ket{\tilde{\Psi}_v}=Q^{\dagger}\ket{\Psi_v}=\sum_{n=\pm}\tilde{\Psi}_v^n(\br)\ket{n}$.
In the above, identity matrix $\mathbbm{1}$ has been eliminated for simplicity, and
\begin{equation}
\begin{aligned}
\bA&=-i\hbar Q^{\dagger}\nabla Q=\sum_{m,n=\pm}\ket{m}\bA^{mn}\bra{n}\\
\bA^{mn}&=-i\hbar\braket{\chi_m|\nabla\chi_n}
\end{aligned},
\end{equation}
where $\ket{\nabla\chi_n}=\nabla\ket{\chi_n}$.
The orthonormality of the internal states guarantee that $i\braket{\chi_m|\nabla\chi_m}$ is real, and $\braket{\nabla\chi_m|\chi_{n\ne m}}=-\braket{\chi_m|\nabla\chi_{n\ne m}}$.\cite{NonAbelianGaugeRMP} 

One can easily identify that $\bA^{mn}$ has the same form as the Berry connection that arises because of the spatial dependence of the internal states $\ket{\chi_{\pm}}$.\cite{GoldmanNonAbelianReview,QianNiuRMP,NonAbelianGaugeRMP} Physically, $H_v^{\text{eff}}$ describes a particle interacting with a non-Abelian gauge potential $\bA$ (i.e. $[A_x,\,A_y]\ne0$) and a scalar potential $-\frac{E_g}{2}+\mathcal{V}_0+\mathcal{V}\sigma_{z}$. Using the expressions in Eq.~(\ref{Eq:InternalDegreeEigenvector}), one can show explicitly that
\begin{equation}
	\bA=
	-\frac{\hbar}{2}\left(\nabla\phi^*\right)
	\begin{pmatrix}
	\cos\zeta&\sin\zeta\\\sin\zeta&-\cos\zeta
	\end{pmatrix}
	+\frac{\hbar}{2}\left(\nabla\zeta\right)\sigma_{y}.\label{Eq:BerryConnectionMatrix}
\end{equation}
It should be noted that $\bA$ is gauge dependent.\cite{NonAbelianGaugeTMDCurvature,GeometricGaugeFieldOpticalLatticeNCooper,GeometricGaugeFieldOpticalLatticeNewJPhys,GoldmanNonAbelianReview,QianNiuRMP,NonAbelianGaugeRMP} For instance, when $\ket{\chi_{\pm}}\rightarrow\ket{\chi_{\pm}}e^{i\alpha_\pm}$, the gauge potential transforms as $\bA^{\pm\pm}\rightarrow\bA^{\pm\pm}+\hbar\nabla\alpha_\pm$ and $\bA^{\pm\mp}\rightarrow\bA^{\pm\mp}e^{-i\left(\alpha_\pm-\alpha_\mp\right)}$. However, physical quantities discussed below are gauge invariant.

One can also define a magnetic field (or Berry curvature) associated with the non-Abelian gauge potential. Within our definition, the canonical momentum reads $\bp+\bA$, thus the covariant derivative is $\textbf{D}=\nabla+\frac{i}{\hbar}\bA$ and the non-Abelian Berry curvature is $\boldsymbol{\mathcal{F}}=\frac{1}{e}\textbf{D}\times\bA=\frac{1}{e}\nabla\times\bA+\frac{i}{e\hbar}[A_x,A_y]$.\cite{NonAbelianGaugeRMP,GoldmanNonAbelianReview,QianNiuRMP} Apart from the usual term that involves the curl of $\bA$, an extra commutator term arises due to the non-Abelian nature of $\bA$. Since $\bA$ is built with a complete basis in the $2\times2$ space, the non-Abelian Berry curvature vanishes.\cite{GoldmanNonAbelianReview,NonAbelianGaugeRMP,QianNiuRMP} One can verify $\boldsymbol{\mathcal{F}}\equiv0$ straightforwardly with Eq.~(\ref{Eq:BerryConnectionMatrix}). However, if the two internal states are well separated in energy (i.e. the separation in $\epsilon_{\pm}$ is much larger than their coupling and kinetic energy of the particles), one can decouple them and consider that the system follows either of them adiabatically. In this adiabatic scenario, an Abelian Berry curvature (pseudo-magnetic field) can be defined with the diagonal elements of the Berry connection in Eq.~(\ref{Eq:BerryConnectionMatrix}) in the familiar way $\bB_{\pm}=\frac{1}{e}\nabla\times\bA^{\pm\pm}$.\cite{NonAbelianGaugeRMP,GoldmanNonAbelianReview} This term can be non-zero. The pseudo-magnetic field discussed in the following refers to $\bB_{\pm}$ specifically, one should not confuse it with the non-Abelain Berry curvature $\boldsymbol{\mathcal{F}}$ or its diagonal elements.

\section{Geometric magnetic field and scalar potential in the adiabatic limit}
Due to the non-Abelian nature of $\bA$, $H_v^{\text{eff}}$ has a matrix form, and the off-diagonal terms represent the coupling between the two internal states. As will be shown later, $\epsilon_{\pm}$ are well separated in a large moir\'e structure (Fig.~\ref{Fig:EandGTwist}). Therefore, when particles move in a large moir\'e structure with small kinetic energies (see more discussions in Sect.~\ref{Sect:ValidityAdiabaticApprox}), one can make an adiabatic approximation by projecting the system onto one of the two internal states using the operator $\hat{P}_{\pm}=\ket{\pm}\bra{\pm}$.\cite{GoldmanNonAbelianReview,NonAbelianGaugeRMP} Here this projection is equivalent to set $\tilde{\Psi}^{\mp}_v=0$, respectively.\cite{NonAbelianGaugeRMP} After doing this, we obtain two separate Schr\"odinger equations
\begin{equation}
	\left[-\frac{1}{2m^{*}}(\bp+\bA^{\pm\pm})^2+\mathcal{G}+E_{\pm}\right]\tilde{\Psi}_v^\pm=\tilde{E}\tilde{\Psi}_v^\pm,
\end{equation}
where $\mathcal{G}=-\frac{1}{2m^*}\bA^{+-}\cdot\bA^{-+}=-\frac{1}{2m^*}|\bA^{+-}|^2$, $E_{\pm}=\mathcal{V}_0\pm\mathcal{V}$, and $\tilde{E}=E+E_g/2$. One can also add negative signs on both sides of the equations to proceed using the language of holes instead of valence band electrons.\cite{HongyiPseudoFieldMoire}

Each equation determines the center-of-mass motion of the system when following one of the internal states adiabatically. The first term describes the kinetic energy of an electron (with charge $-e$) in the presence of a geometric magnetic field $\bB_{\pm}=\frac{1}{e}\nabla\times\bA^{\pm\pm}$. The second term $\mathcal{G}$ is the so-called geometric scalar potential, representing the (negative) kinetic energy associated with the micromotion due to the force originates from particle's virtual transition between the two internal states.\cite{NonAbelianGaugeRMP,GoldmanNonAbelianReview,GeometricScalarPotentialMeaningEPL2008} They are goemetric because they depend on the spatial variations of the spherical angles $(\zeta,\,\phi^*)$ as will be shown explicitly in Eqs.~(\ref{Eq:B+_rigid_twist}) and (\ref{Eq:G_rigid_twisting}). As $\bA^{\pm\pm}$ and $\bB_{\pm}$ are vectors, and $\mathcal{G}$ is a scalar, they are Abelian in nature. It should be pointed out that non-zero $\bB_{\pm}$ and $\mathcal{G}$ arise because of the adiabatic elimination of the other internal state.\cite{GoldmanNonAbelianReview} In the following, we will look at how the moir\'e magnetic field and scalar potentials behave explicitly.

\subsection{Magnetic field with non-zero quantized flux}
In this section we will look at the moir\'e magnetic field. With the gauge choice in Eq.~(\ref{Eq:InternalDegreeEigenvector}), one obtains $\bA^{++}=-\bA^{--}$ from Eq.~(\ref{Eq:BerryConnectionMatrix}). Thus it is obvious that $\bB_{+}=-\bB_{-}$ and we will only consider $\bB_{+}=B_{+}\hat{z}$. Straightforward calculations yield\cite{NonAbelianGaugeRMP,NonAbelianGaugeTMDCurvature}
\begin{equation}
\begin{aligned}
\bB_{+}
&=-\frac{1}{4\pi}\Phi_0\nabla\left(\cos\zeta\right)\times\nabla\phi^{*}\\
&=-\frac{1}{4\pi}\Phi_0\nabla\braket{\sigma_z}_+\times\nabla\left(\arg\braket{\bsigma}_+\right)
\end{aligned},\label{Eq:B+_rigid_twist}
\end{equation}
where $\Phi_0=h/e$ is the magnetic flux quantum. It is clear that the pseudo-spin distribution is crucial for the emergence of the magnetic field: Non-zero magnetic field will emerge only if the pseudo-spin varies in space as well as the out-of-plane pseudo-spin and the in-plane pseudo-spin orientation exhibit noncollinear gradients.

Before proceeding further, one may notice that $\bA^{++}=-\frac{\hbar}{2}\left(\nabla\phi^*\right)\cos\zeta$, which is used to derive the magnetic field, exhibits singularities.\cite{GeometricGaugeFieldOpticalLatticeNCooper,GeometricGaugeFieldOpticalLatticeNewJPhys,SpielmanFluxLatticeComment} For instance, around the north and south poles on the Bloch sphere, $\bA^{++}\rightarrow\mp\frac{\hbar}{2}\left(\nabla\phi^*\right)$, which are ill-defined because $\nabla\phi^*$ yields different values from distinct directions. This is because $\phi^*$ in $\ket{\chi_{+}}$ of Eq.~(\ref{Eq:InternalDegreeEigenvector}) is not well-defined for $\zeta=0$ and $\pi$. The line going through $\zeta=0$ and $\pi$ is the so-called Dirac string, which acts as the solenoid in the  Aharonov-Bohm effect.\cite{GeometricGaugeFieldOpticalLatticeNCooper,GeometricGaugeFieldOpticalLatticeNewJPhys,SpielmanFluxLatticeComment} The location of the Dirac string is gauge dependent and not observable. For instance, when $\ket{\chi_{+}}\rightarrow\ket{\chi_{+}}e^{i\frac{\phi^{*}}{2}}$, the gauge potential becomes $\bA^{++}=-\frac{\hbar}{2}\left(\nabla\phi^*\right)\left(\cos\zeta-1\right)$, which is ill-defined only at the south pole, rendering a Dirac string on the semi-infinite negative $z$-axis. Nonetheless, the existence of the Dirac string is gauge invariant, and it is a necessary condition to have non-zero quantized magnetic flux as will be shown in the following.\cite{GeometricGaugeFieldOpticalLatticeNCooper,GeometricGaugeFieldOpticalLatticeNewJPhys,SpielmanFluxLatticeComment}

Despite $\bA^{++}$ exhibit singularities, the magnetic field is smooth. One can also express $\bB_{+}$ in terms of $\vec{n}$, which is smooth and well defined everywhere on the Bloch sphere: $\bB_{+}=\frac{1}{4\pi}\Phi_0\vec{n}\cdot\left(\partial_{x}\vec{n}\times\partial_{y}\vec{n}\right)\hat{z}$ or $\bB_{+}=\frac{1}{4\pi}\Phi_0\left(\nabla n_x\times\nabla n_y\right)/n_z$.\cite{GeometricGaugeFieldOpticalLatticeNCooper,GeometricGaugeFieldOpticalLatticeNewJPhys}
Using Eq.~(\ref{Eq:B+_rigid_twist}), one can find that the magnetic flux reads
\begin{equation}
\begin{aligned}
\Phi
&=\frac{1}{4\pi}\Phi_0\int_{\text{real}} \left(\sin\zeta\nabla\zeta\times\nabla\phi^{*}\right)\cdot\hat{z}\,dxdy\\
&=\frac{1}{4\pi}\Phi_0\int_{\text{Bloch}} \sin\zeta\,d\zeta d\phi^{*}\\
&=\frac{1}{4\pi}\Phi_0\Omega
\end{aligned},\label{Eq:FluxInTermsOfSolidAngle}
\end{equation}
where $\Omega=\int \sin\zeta\,d\zeta d\phi^{*}$ is the solid angle covered by the pseudo-spin trajectory on the Bloch sphere. The first integral is carried out over an area in real space, while the second integral is over the corresponding region on the Bloch sphere. To arrive at the second line, we used the fact that $\left(\nabla\zeta\times\nabla\phi^{*}\right)\cdot\hat{z}$ is the Jacobian determinant relating Cartesian to spherical coordinates. Therefore, the magnetic flux characterizes how the pseudo-spin rotates on the Bloch sphere when a particle moves in the real space. Especially, it will be quantized as $\Phi=N\Phi_0$ if the pseudo-spin winds integer $N$ times on the Bloch sphere. It shows that the geometric phase gained by the pseudo-spin ($\Omega/2$) can be interpreted as arising from the geometric  magnetic field.

\begin{figure}[ht]
	\includegraphics[width=3.4in]{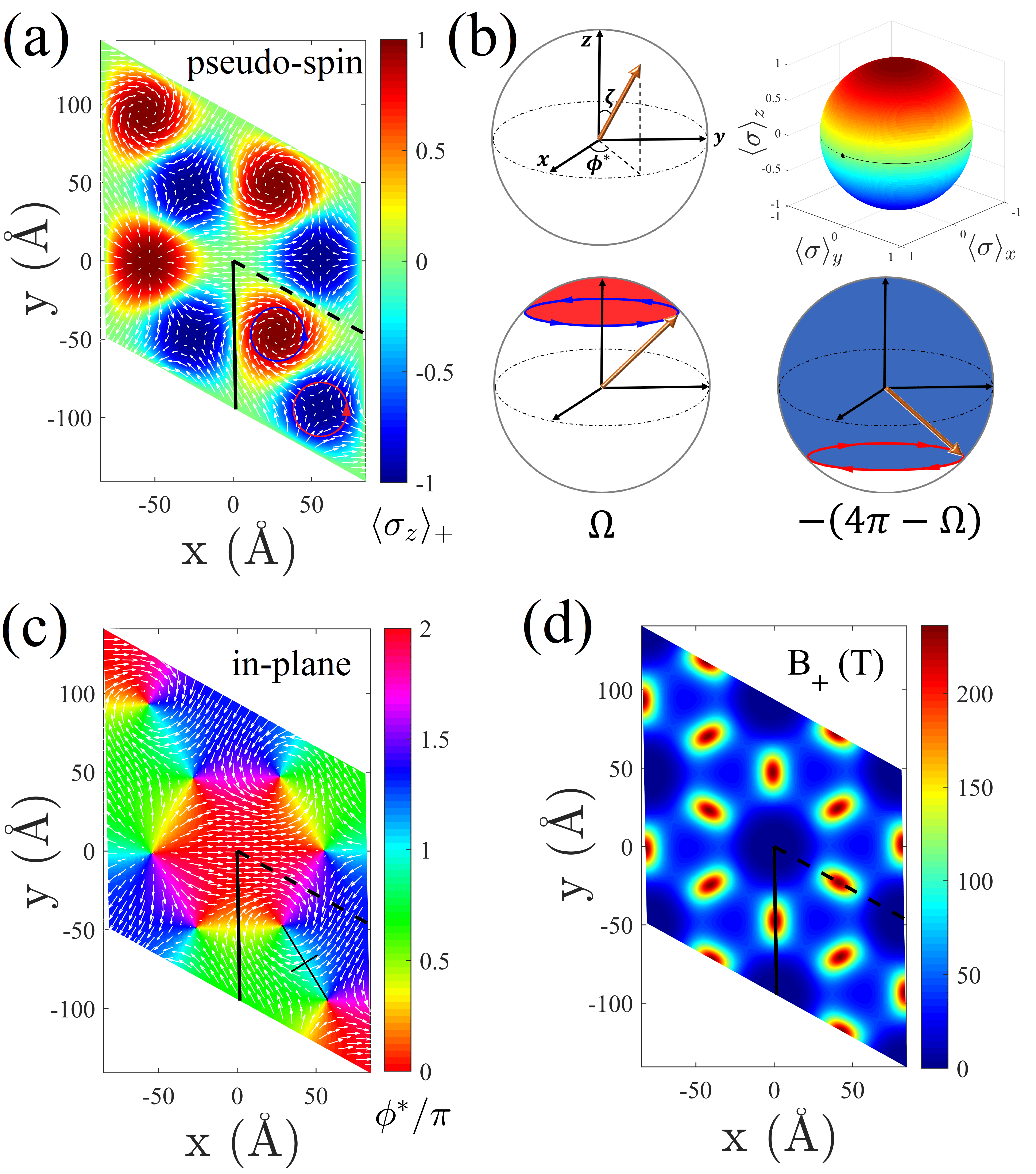}
	\caption{Pseudo-spin distribution and magnetic field for a $2^\circ$ twisted bilayer. (a) The arrows and background color represent in-place and out-of-plane pseudo-spin, respectively. The solid and dashed lines denote the moir\'e primitive vectors $\bL_1$ and $\bL_2$. The circles are two real space trajectories for a moving particle. (b) The top left panel shows schematics of the pseudo-spin $\braket{\vec{\sigma}}_+$ (brown arrow) and its polar and azimuthal angles. The top right panel shows the surface defined by the end points of all the pseudo-spin vectors $\braket{\vec{\sigma}}_+$ in the moir\'e unit cell, which form a Bloch sphere. The solid and dashed curves correspond to the solid and dashed lines in panel (a), and the black dot corresponds to the origin. The bottom left/right panel shows the trajectory of the pseudo-spin on the Bloch sphere when the particle moves in the blue/red circle in real space in panel (a). The resultant solid angle is $\Omega$ and $-(4\pi-\Omega)$, respectively. (c) The arrows and background color represent in-place pseudo-spin and its orientation $\phi^{*}$, respectively. (d) Spatial distribution of $B_{+}$.}
	\label{Fig:PseudospinAndMagneticField2deg}
\end{figure}

Let us now look at the pseudo-spin distribution $\braket{\vec{\sigma}}_{+}$ before discussing features of the magnetic field.
Fig.~\ref{Fig:PseudospinAndMagneticField2deg}(a) shows $\braket{\vec{\sigma}}_{+}$ for a $2^\circ$ twisted bilayer MoSe$_2$ within four moir\'e unit cells (the solid and dashed lines represent the edges of one unit cell).
\footnote{By inspecting their distributions in the six $R^X_M$ and $R^M_X$ locals surrounding the origin, it appears that equivalence among identical local crystalline structures is broken (for instance, arrows in the three red regions follow distinct patterns). Recall that in-plane pseudo-spin follows the phase of $\tilde{U}^{*}_{vv}$, one finds that this breaking of symmetry in in-plane pseudo-spin orientation is caused by the transformation $\tilde{U}_{vv}=U_{vv}e^{i\frac{\theta}{2}}e^{i\left(\bK-\tilde{\bK}\right)\cdot\br}$ (also see Appendix~\ref{App:FourBandContinuumModel}), where $\tilde{U}_{vv}$ differs from $U_{vv}$ by a space-dependent phase $\left(\bK-\tilde{\bK}\right)\cdot\br$. One can verify that the phase of $U_{vv}$ exhibits the desired translation symmetry, i.e. the three red (blue) regions are equivalent, while $\left(\bK-\tilde{\bK}\right)\cdot\br$ destroys it in $U_{vv}^{*}$ (also see discussions in C. Bena and G. Montambaux, New J. Phys. 11, 095003 (2009)). However, physical quantities, e.g. the magnetic field, are not affected and exhibit the correct symmetries of the system.}
The top right panel of Fig.~\ref{Fig:PseudospinAndMagneticField2deg}(b) presents the distribution of the pseudo-spin in spherical coordinates, which shows that the pseudo-spin forms a closed surface if all the points in the moir\'e unit cell were visited.
From panel (a) one can see that $\braket{\sigma_z}_+$ (represented by the background color) is maximum but exhibits opposite signs around $R^X_M$ and $R^M_X$ local stackings, which is related to the symmetry properties of $\mathcal{V}_z$. Meanwhile, in-plane pseudo-spin (denoted by the arrows) exhibits vortex and anti-vortex structures.\cite{WuMacDonaldPRL2019} Therefore, the pseudo-spin forms a skyrmion lattice.\cite{NagaosaSkyrmionNatNano2013} The vortex/anti-vortex texture is the origin of the presence of singularities in $\bA^{++}$.

The above observations imply that the magnetic flux is non-zero and quantized.\cite{GeometricGaugeFieldOpticalLatticeNCooper,GeometricGaugeFieldOpticalLatticeNewJPhys,SpielmanFluxLatticeComment} Consider the two counterclockwise loops in the moir\'e unit cell in Fig.~\ref{Fig:PseudospinAndMagneticField2deg}(a), which represent two real space trajectories for a moving particle. Their corresponding routes on the Bloch sphere are shown schematically in the lower panels of Fig.~\ref{Fig:PseudospinAndMagneticField2deg}(b). As the vorticity of the in-plane pseudo-spin are opposite in the two loops, the pseudo-spin rotates counterclockwise (blue) and clockwise (red), respectively. If the solid angle covered by the red area enclosed by the blue loop is $\Omega$, the red loop will correspond to a $-(4\pi-\Omega)$ solid angle because the surface normal is opposite. In terms of the geometric phase acquired by the pseudo-spin, the two loops will contribute equally because a phase of $-2\pi+\Omega/2$ is equivalent to $\Omega/2$.\cite{SpielmanFluxLatticeComment} Overall, the magnetic flux through one moir\'e unit cell should equal to one flux quantum due to the $4\pi$ solid angle of the Bloch sphere. In fact, when the pseudo-spin exhibits skyrmion type structures, each half of the unit cell contributes exactly $\pm1/2$ flux quantum, the sign should be determined by the product of the signs of $\braket{\sigma_{z}}_+$ and vorticity of $\braket{\bsigma}_+$ ($+$ in the current situation).\cite{GeometricGaugeFieldOpticalLatticeNCooper}
Lattices with such non-zero quantized flux per unit cell was proposed previously based on elaborate optical schemes. \cite{GeometricGaugeFieldOpticalLatticeNCooper,GeometricGaugeFieldOpticalLatticeNewJPhys,SpielmanFluxLatticeComment} Here one can see that twisted bilayer TMD is a natural platform to realize such flux lattices.

Fig.~\ref{Fig:PseudospinAndMagneticField2deg}(c) shows the in-plane pseudo-spin distribution and its direction $\arg \braket{\bsigma}_+=\phi^{*}$ as the background color. The longer black line inside the unit cell connects the $R^X_M$ and $R^M_X$ local stackings, where one may notice that $\braket{\sigma_{z}}_+$ varies the greatest along this direction (see Fig.~\ref{Fig:PseudospinAndMagneticField2deg}(a)). In the perpendicular direction, as specified by the shorter black line, $\phi^{*}$ has the largest variation. As shown in Eq.~(\ref{Eq:B+_rigid_twist}), the magnetic field is proportional to the cross product of the gradients of $\braket{\sigma_z}_+$ and $\phi^{*}$. Therefore, the intersection of these two directions is expected to determine the location of the maximum magnetic field.
Fig.~\ref{Fig:PseudospinAndMagneticField2deg}(d) shows the distribution of $B_{+}$. As expected, magnetic field exhibits six-fold rotational symmetry with hot spots in the junction between $R^M_X$ and $R^X_M$ stackings (around the intersection of the two lines in Fig.~\ref{Fig:PseudospinAndMagneticField2deg}(c)). The field is non-negative everywhere, so the magnetic flux must be non-zero. We have confirmed numerically that each moir\'e unit cell encloses exactly one flux quantum, consistent with the analysis based on solid angle coverage on the Bloch sphere.

As varying the twist angle leaves the profile of the pseudo-spin texture (thus the Bloch sphere) unaffected, the magnetic flux will remain quantized independent of the size of the moir\'e tuned by $\theta$.
Such a constant magnetic flux obviously can be employed to adjust the magnitude of the magnetic field by tuning the size of the moir\'e. For a moir\'e pattern with period around $10$ nm, the magnetic field can reach the order of $100$ T (Fig.~\ref{Fig:PseudospinAndMagneticField2deg}(d)). The area of the moir\'e unit cell is $S_{\text{moir\'e}}(\theta)=|\bL_1\times\bL_2|\approx\frac{\sqrt{3}a^2}{2\theta^2}$. Therefore, changing the twist angle from e.g. $2^\circ$ to $0.5^\circ$ increases $S_{\text{moir\'e}}$ by more than a factor of $10$, the average magnetic field will decrease by one order of magnitude accordingly.

To close the discussions in this section, we will comment on the results of $\frac{1}{e}\oint \bA^{++}\cdot d\bl$, which is often employed to evaluate the magnetic flux using Stokes' theorem. If the integral is performed on the boundaries of a moir\'e unit cell, the result will vanish due to the periodicity of the system. This clearly means that Stokes' theorem is invalid here, as the moir\'e unit cell is not simply connected because of the singularities in $\bA^{++}$. However, one can show that the results obtained by performing the integral on infinitesimal loops around the singularities will yield the correct magnetic flux (this is equivalent to apply Stokes' theorem after excluding the Dirac string).\cite{GeometricGaugeFieldOpticalLatticeNewJPhys} To facilitate discussions, let us choose the gauge such that $\bA^{++}=-\frac{\hbar}{2}\left(\nabla\phi^*\right)\left(\cos\zeta-1\right)$, which has singularity only at the south pole. If we choose an infinitesimally small loop around the south pole, which corresponds to a small loop around the center of the blue region in real space in Fig.~\ref{Fig:PseudospinAndMagneticField2deg}(a), then $\frac{1}{e}\oint_{r\rightarrow0} \bA^{++}\cdot d\bl\rightarrow\frac{\hbar}{e}\oint \nabla\phi^*\cdot d\bl=\frac{h}{e}$, i.e. one flux quantum. The infinitesimal loop splits the Bloch sphere into two domains: the northern domain $D_\uparrow$ covering almost the entire sphere (with $\bA^{++}$ being well defined), and the southern domain $D_\downarrow$ with vanishing area ($\bA^{++}$ is singular at the south pole). With $\bB_{+}$ being a smooth function, the magnetic flux from the southern domain vanishes $\int_{D_\downarrow}\bB_{+}\cdot d\bS\rightarrow0$, hence the flux from the northern sphere is one flux quantum $\int_{D_\uparrow}\bB_{+}\cdot d\bS\rightarrow\int_{D_\uparrow+D_\downarrow}\bB_{+}\cdot d\bS=\frac{h}{e}$. Therefore, $\frac{1}{e}\oint_{r\rightarrow0} \bA^{++}\cdot d\br=\int\bB_{+}\cdot d\bS$ inside the moir\'e unit cell if the the Dirac string at the south pole is excluded. The magnetic flux will be zero if $\bA^{++}$ does not have any singularity (as Stokes' theorem will be valid in the entire moir\'e unit cell and the integral of $\bA^{++}$ vanishes along the moir\'e boundaries), showing the importance role of the vortex/anti-vortex texture of the pseudo-spin.

\subsection{Scalar potentials and twist dependent effective tight-binding lattices}
In this section we will discuss the properties of the scalar potentials $E_\pm$ and $\mathcal{G}$ in a moir\'e lattice. We first consider $E_{\pm}=\mathcal{V}_0\pm\mathcal{V}$, whose magnitudes are independent of the moir\'e size as explained in the following. Recall that $E_{\pm}$ are eigenvalues of $\mathcal{U}_{v}$ in Eq.~(\ref{Eq:ValenceBandMoirePotential}), which is a function of $\bG_i\cdot\br$. This makes $E_{\pm}$ functions of $\bG_i\cdot\br$ as well, i.e. $E_{\pm}=E_{\pm}(\bG_1\cdot\br,\,\bG_2\cdot\br)$. Different twist angles will yield moir\'e patterns with distinct $\bG_i(\theta)$ and $\bL_i(\theta)$. If we parametrize $\br$ in a moir\'e unit cell as $\br_{n_1n_2}(\theta)=n_1\bL_1(\theta)+n_2\bL_2(\theta)$, where $n_1,\,n_2\in[0,\,1]$ are continuous variables, one can obtain $\bG_i(\theta)\cdot\br_{n_1n_2}(\theta)=2\pi n_i$, which is independent of $\theta$. Therefore, equivalent crystalline locations (i.e. shared $n_1$ and $n_2$) in moir\'e structures with distinct $\theta$ exhibit identical $E_{\pm}(\br_{n_1n_2}(\theta))\equiv E_{\pm}(2\pi n_1,\,2\pi n_2)$. 

\begin{figure}[ht]
	\includegraphics[width=3.4in]{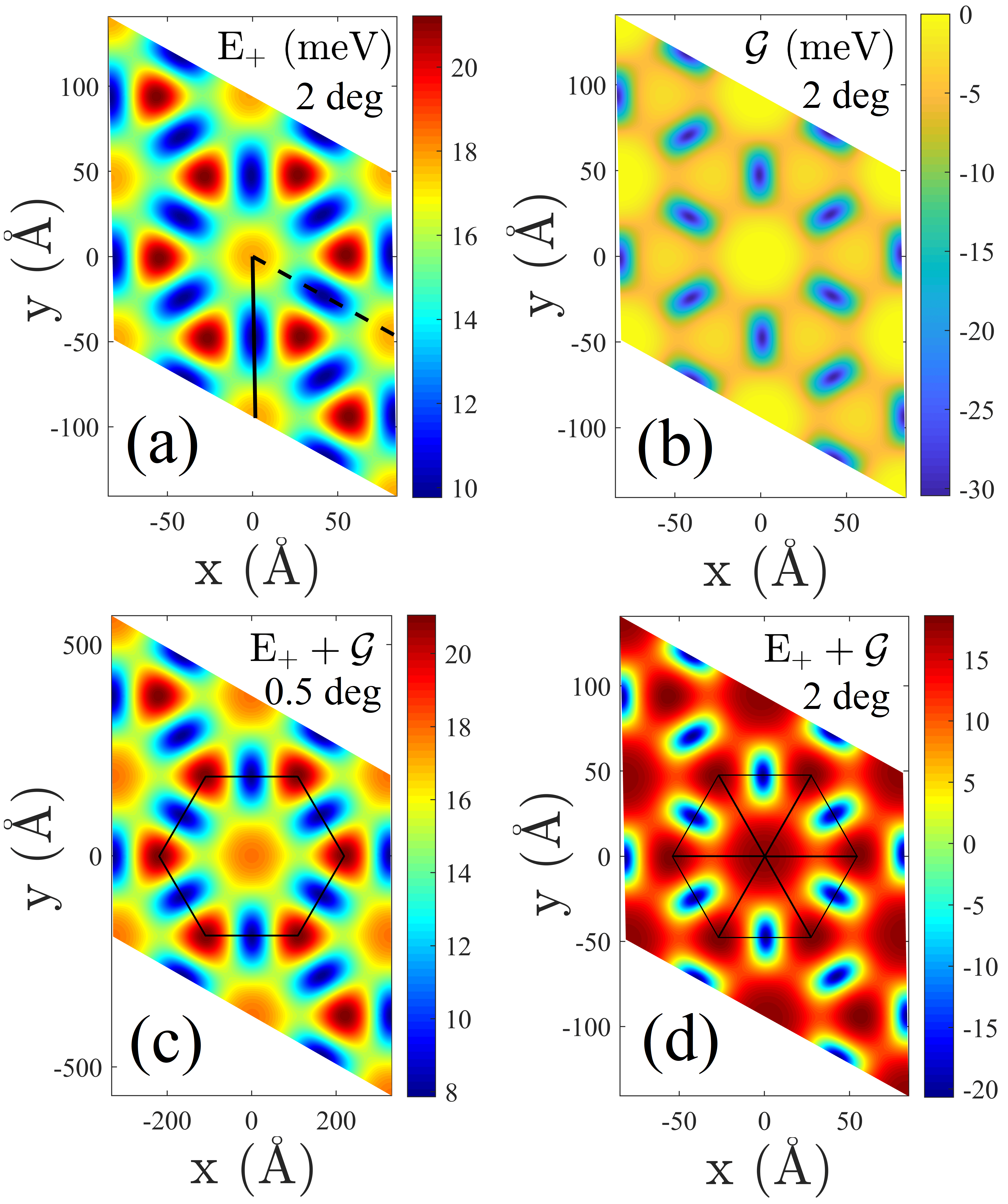}
	\caption{Spatial distribution of scalar potentials. (a) $E_{+}$, (b) $\mathcal{G}$, and (d) $\mathcal{G}+E_{+}$ in four moir\'e unit cells for a $2^\circ$ twisted bilayer. (c) $\mathcal{G}+E_{+}$ for a $0.5^\circ$ twisted bilayer. Black solid (dashed) line in panel (a) denotes $\bL_1$ ($\bL_2$). Triangles and hexagon in panels (c) and (d) sketch the lattice structures formed by the dark red spots.}
	\label{Fig:EandGTwist}
\end{figure}

Fig.~\ref{Fig:EandGTwist}(a) shows the spatial dependence of $E_{+}$ in four moir\'e unit cells for a $2^\circ$ twisted MoSe$_2$ bilayer. $E_-$ exhibits almost the same profile as $E_+$ in the negative $E$ axis with slight asymmetry in magnitude caused by $\mathcal{V}_0$, which shifts the mid-gap position. Clearly, $E_+$ is strong around $R^X_M$ and $R^M_X$ stackings, while it is the weakest in regions where the magnetic field is large (compare with Fig.~\ref{Fig:PseudospinAndMagneticField2deg}(d)).

Now let us look at the more interesting geometric scalar potential $\mathcal{G}$. Straightforward calculations yield\cite{NonAbelianGaugeRMP}
\begin{equation}
	\mathcal{G}=-\frac{\hbar^2}{8m^*}\left[\left(\nabla\zeta\right)^2+\sin^2\zeta\left(\nabla\phi^{*}\right)^2\right].\label{Eq:G_rigid_twisting}
\end{equation}
Consequently, $\mathcal{G}$ depends on gradients of the spherical angles of pseudo-spin, as well as magnitude of the in-plane pseudo-spin ($\sin\zeta$). $\mathcal{G}$ can also be written as $\mathcal{G}=-\frac{1}{2m^*}\left(p_{\zeta}^2+p_{\phi^*}^2\right)$, where $\bp_{\zeta}=\frac{\hbar}{2}\nabla\zeta$ and $\bp_{\phi^*}=\frac{\hbar}{2}\sin\zeta\left(\nabla\phi^*\right)$ can be understood as the micromotion momentum along $\zeta$ and $\phi^*$ direction, respectively. Written in this way, it is clearer that $\mathcal{G}$ represents the micromotion kinetic energy due to forces applied upon the particles when virtual transition between internal states occur.\cite{NonAbelianGaugeRMP,GoldmanNonAbelianReview,GeometricScalarPotentialMeaningEPL2008}
Fig.~\ref{Fig:EandGTwist}(b) shows the typical results of $\mathcal{G}$. One may notice that $\mathcal{G}$ and the magnetic field share significant similarities: (i) The profile of $\mathcal{G}$ resembles that of the magnetic field in Fig.~\ref{Fig:PseudospinAndMagneticField2deg}(d). (ii) The `flux' of $\mathcal{G}$ through the moir\'e unit cell is independent of $\theta$ as well. The latter implies that $\mathcal{G}$ decreases with the increase of moir\'e size and becomes negligible compared to $E_+$ at small twist angles.\cite{HongyiPseudoFieldMoire} Figs.~\ref{Fig:EandGTwist}(c) and (d) show results of $E_++\mathcal{G}$ in the case of $\theta=0.5^\circ$ and $2^\circ$, respectively. One can clearly see that Fig.~\ref{Fig:EandGTwist}(c) resembles Fig.~\ref{Fig:EandGTwist}(a) (their spatial scales are very different though) because $\mathcal{G}\ll E_+$ and can be neglected when $\theta=0.5^\circ$. 

It is also interesting to notice that the dark red spots in Fig.~\ref{Fig:EandGTwist}(c), which act as trapping sites for \textit{holes}, form a honeycomb lattice structure as indicated by the hexagon. These trapping sites serve as the lattice sites for the effective tight-binding description of the moir\'e lattice, which has been proposed recently in Ref.~\cite{WuMacDonaldPRL2019}. In contrast, the effective tight-binding lattice becomes `decorated triangular' \cite{DecoratedTriangularLattice,DecoratedTriangularLatticeTopologyChange} if $\theta$ becomes larger as shown in Fig.~\ref{Fig:EandGTwist}(d), where both the center and corners of the hexagon become trapping sites. This is because $\mathcal{G}$ now contributes a larger negative weight, so the magnitude of the potential at the hexagonal corners decreases and becomes comparable to that in the center. A three-orbital effective tight-binding model based on such lattice structure was proposed in Ref.~\cite{HongyiPseudoFieldMoire} to describe twisted homobilayer TMD.  If the twist angle is increased further, the trapping site at the center of the hexagon dominates and the corners can be neglected, the scalar potential then forms a simple triangular lattice.\cite{DecoratedTriangularLatticeTopologyChange} Such lattice structure transitions, when combined with the magnetic field background, may cause changes in the electronic and topological properties of moir\'e lattices.

\begin{figure}[ht]
	\includegraphics[width=3.4in]{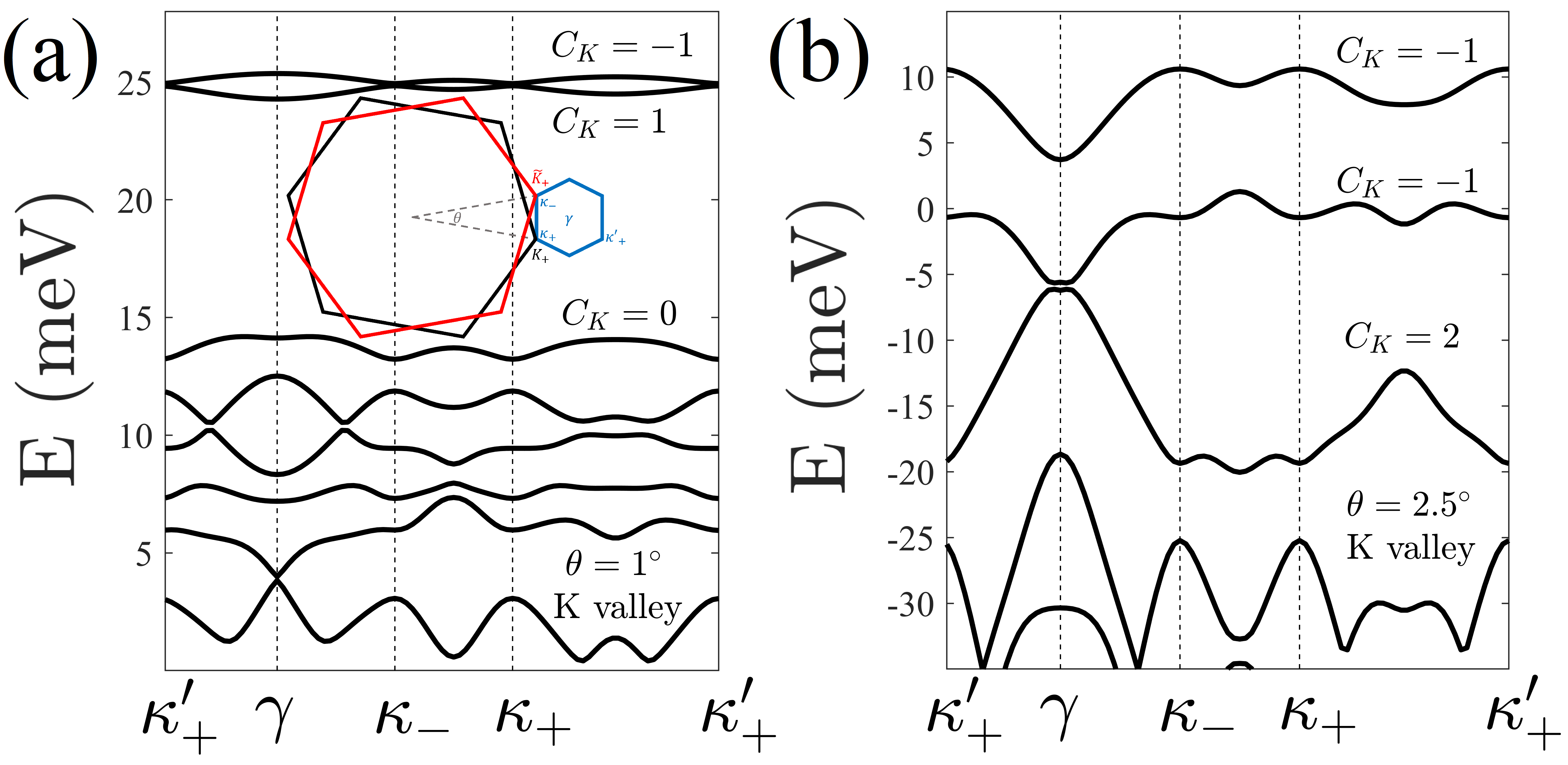}
	\caption{Comparison of moir\'e mini bands and Chern numbers of twisted bilayer MoSe$_2$ with different twist angles. Results shown are obtained from the $K$ valley (with spin down) of the monolayers, while the ones from the $-K$ valley (with spin up) are their time reversal. (a) $\theta=1^\circ$. There is a tiny gap between the two topmost bands, which is unobservable in the current scale. (b) $\theta=2.5^\circ$. Inset in (a) shows the monolayer Brillouin zone (black and red hexagons) and moir\'e mini Brillouin zone (blue hexagon).}
	\label{Fig:BrillouinZoneAndBandStructureTwist}
\end{figure}

Fig.~\ref{Fig:BrillouinZoneAndBandStructureTwist} shows the moir\'e mini bands obtained from Eq.~(\ref{Eq:OriginalHamiltonian}) at two different twist angles, as examples to illustrate the effects of the moir\'e pseudo-magnetic field and geometric scalar potential. Chern numbers for the three topmost bands are presented as well. First, one may notice that the bands are shifted downward in energy when the twist angle $\theta$ is increased (note the different vertical scales of the two panels). Such an energy shift can be attributed to the variation of scalar potential with $\theta$ (c.f. Fig.~\ref{Fig:EandGTwist}(c, d)). Second, one can find that the topmost bands exhibit nontrivial Chern numbers.\cite{WuMacDonaldPRL2019,HongyiPseudoFieldMoire} This is consistent with the presence of moir\'e pseudo-magnetic field, which realizes fluxed lattices underlying the quantum spin Hall effect.\cite{HaldaneQHE} Furthermore, it is observed that the band width increases with $\theta$, some bands eventually cross (e.g. the 2nd and 3rd in Fig.~\ref{Fig:BrillouinZoneAndBandStructureTwist}(b)) and their Chern numbers are modified. We believe that such changes are caused by the complex interplay of the scalar potential and the underlying pseudo-magnetic field. Changes in the landscape of the scalar potential as well as moir\'e period while tuning $\theta$ affects the effective tight-binding description of the moir\'e lattice (i.e. lattice geometry, hence magnitude and phase of the hopping).\cite{HongyiPseudoFieldMoire,ObtainHoppingParameterPRA2014}

\subsection{Validity of the adiabatic approximation}\label{Sect:ValidityAdiabaticApprox}
For the adiabatic approximation to be valid, it is desirable to tune the kinetic energy of particles below the energy spacing of the internal states.\cite{GeometricGaugeFieldOpticalLatticeNCooper} The off-diagonal terms of $\left(\bp+\bA\right)^2$ will cause mixing between the two internal states. Therefore, energy associated with its off-diagonal terms, which read $\bp\cdot\bA^{\pm\mp}+\bA^{\pm\mp}\cdot\bp$ with the gauge choice in Eq.~(\ref{Eq:InternalDegreeEigenvector}), also needs to be small compared to the energy gap between the two internal states. As is discussed previously, spatial dependence of the moir\'e potential is expressed in terms of $\bG_i\cdot\br$. Therefore, the coupling energy can be estimated as $\frac{\hbar^2G^2_i}{2m^*}\approx\frac{8\pi^2\hbar^2}{3m^*L^2}=\frac{16\pi^2}{3E_g}\left(\frac{\hbar v_F}{L}\right)^2$, where $L$ is the moir\'e period (Table~\ref{Table}). For a MoSe$_2$ moir\'e with $L=10$ nm, this corresponds to an energy about $35$ meV, which is close to the gap size of the two internal states. Therefore, adiabatic approximation works well in the low energy and large moir\'e limit.

\section{Tunability of the properties of moir\'e}
We have seen that twist angle can be used to tune certain properties of the moir\'e lattice. In this section, we will explore utilizing external means, i.e. interlayer bias and uniform strain, to tailor the properties of moir\'e.

\subsection{Interlayer bias tuning}\label{Sect:BiasEngineering}

\begin{figure*}[ht]
	\includegraphics[width=6in]{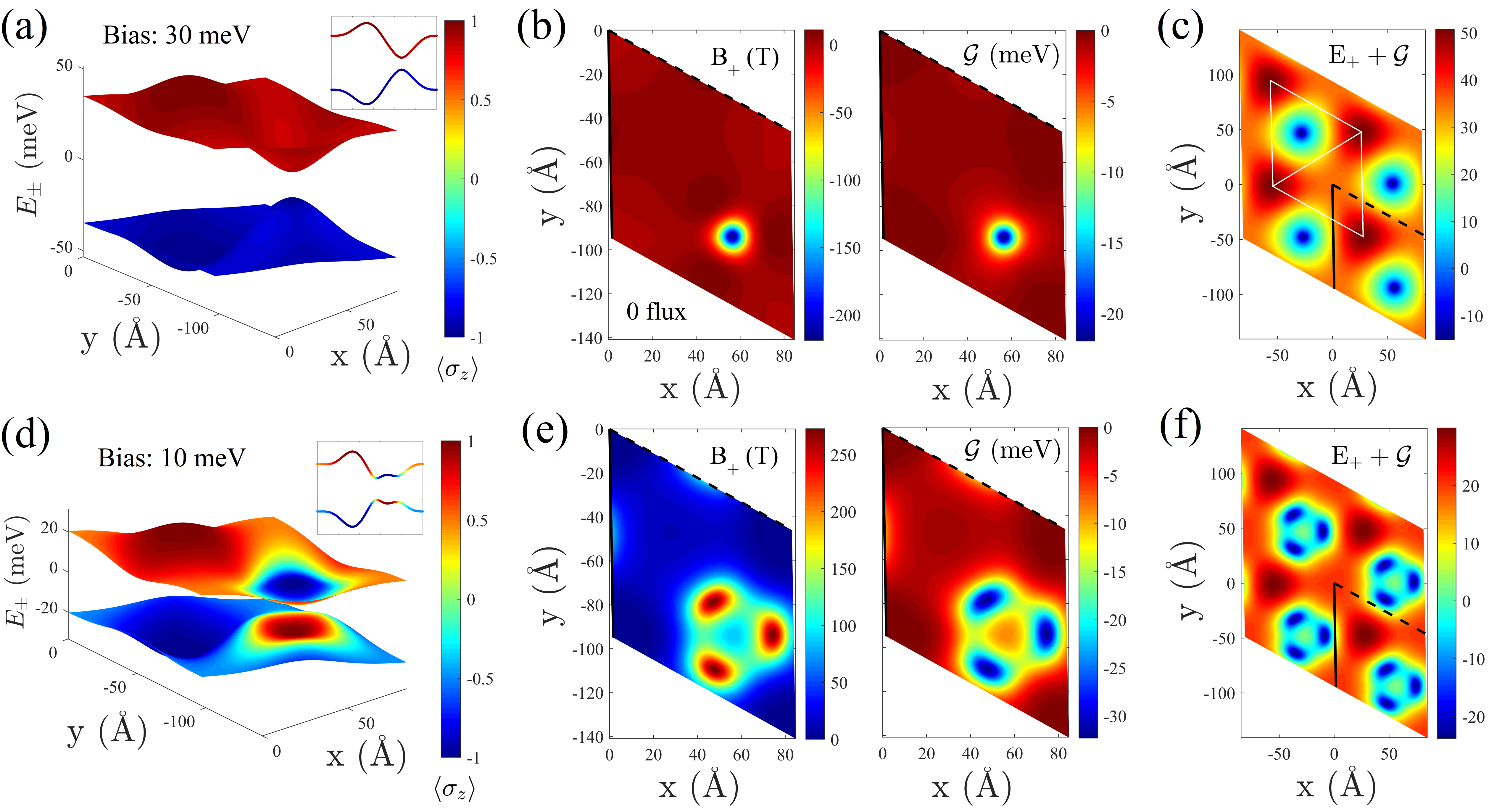}
	\caption{$E_{\pm}$, $B_{+}$, $\mathcal{G}$ and $E_+ + \mathcal{G}$ \textit{vs} interlayer bias for a $2^\circ$ twisted bilayer. (a--c) $V_{\mathcal{E}}=30$ meV $>|V_c|$, (d--f) $V_{\mathcal{E}}=10$ meV $<|V_c|$. Color coding in panels (a, d) represents the out-of-plane pseudo-spin $\braket{\sigma_{z}}_\pm$. Insets show the lines cuts along the diagonal of the moir\'e unit cell. Note that panels (c, f) show results in four moir\'e unit cells, while the rest show results in one unit cell. White triangles in panel (c) sketch the triangular lattice formed by the dark red spots. Here $|V_c|=22.3$ meV for twisted bilayer MoSe$_2$.}
	\label{Fig:E_and_B_G_vs_bias}
\end{figure*}

As $\braket{\sigma_{z}}_\pm\propto\mathcal{V}_z$, interlayer bias can be employed to tune the out-of-plane pseudo-spin, thus the properties of moir\'e. An interlayer bias places the two layers at an additional potential $\pm V_\mathcal{E}$ respectively, where $V_\mathcal{E}=\frac{1}{2}e\mathcal{E}d$ with $\mathcal{E}$ the perpendicular electric field, and $d$ the interlayer separation. Therefore, $\mathcal{V}_z$ should be replaced by $\mathcal{V}_{z,\mathcal{E}}=\mathcal{V}_z+V_\mathcal{E}$. 
The most prominent change occurs when $V_{\mathcal{E}}$ is tuned to a critical value (say $V_{c}$) such that $\mathcal{V}_{z,\mathcal{E}}$ and $\tilde{U}_{vv}$ simultaneously vanish at $(\bL_1+\bL_2)/3$ or $2(\bL_1+\bL_2)/3$ (center of $R_M^X$ or $R_X^M$ local stackings).
In such a situation, $\epsilon_\pm$ (as well as $E_\pm$) become degenerate with the energy gap between them closes at the said locations. It is found that the magnetic flux is one flux quantum if $V_{\mathcal{E}}<|V_c|$, while it vanishes when $V_{\mathcal{E}}>|V_c|$ (see Fig.~\ref{Fig:E_and_B_G_vs_bias}).\cite{HongyiPseudoFieldMoire} 

Such behavior can be understood as the following: The interlayer bias acts like another staggered potential in $\mathcal{U}_{\text{ps}}$, where $\mathcal{V}_z$ and $V_{\mathcal{E}}$ compete. When $V_{\mathcal{E}}>|V_c|$, the two bands $E_{\pm}$ (or equivalently $\epsilon_{\pm}$) are far apart and stay gapped throughout the moir\'e lattice. The out-of-plane pseudo-spin distribution $\braket{\sigma_z}_\pm$ for each internal state has a fixed sign independent of location in Fig.~\ref{Fig:E_and_B_G_vs_bias}(a), while $\braket{\bsigma}_\pm$ is unaffected by the bias (same as that in Fig.~\ref{Fig:PseudospinAndMagneticField2deg}(a)). This means that the pseudo-spin does not enclose a closed surface, e.g. $\braket{\vec{\sigma}}_+$ always stays on the northern semi-sphere. Meanwhile, in-plane pseudo-spin texture indicates that $\braket{\vec{\sigma}}_+$ rotates clockwise then counterclockwise with vanishing net solid angle. As mentioned before, one can also arrive at the same conclusion by checking the sign of $\braket{\sigma_{z}}_+$ and the vorticity of $\braket{\bsigma}_+$. For instance, $\text{sgn}\braket{\sigma_{z}}_+\equiv+$ in Fig.~\ref{Fig:E_and_B_G_vs_bias}(a), while the vorticity of $\braket{\bsigma}_+$ exhibits opposite signs on the two equilateral halves of the unit cell (Fig.~\ref{Fig:PseudospinAndMagneticField2deg}(a)). Therefore, the magnetic flux from each half unit cell is $\pm1/2$ flux quantum respectively, and they cancel each other. When $V_{\mathcal{E}}$ decreases, $E_+$ and $E_-$ approach each other, eventually they touch at $(\bL_1+\bL_2)/3$ or $2(\bL_1+\bL_2)/3$ and become gapless when $V_{\mathcal{E}}=|V_c|$. The gap reopens with the occurrence of band inversion, i.e. $\braket{\sigma_z}_\pm$ flips sign near the gap, when $V_{\mathcal{E}}$ is decreased further to $V_{\mathcal{E}}<|V_c|$ in Fig.~\ref{Fig:E_and_B_G_vs_bias}(d). $\braket{\sigma_z}_\pm$ changing sign within the moir\'e unit cell yields a closed Bloch surface.
Such topological transition in the layer pseudo-spin texture results in the quantized jump of magnetic flux from zero to one flux quantum.\cite{HongyiPseudoFieldMoire}

Fig.~\ref{Fig:E_and_B_G_vs_bias} shows the results of $E_{\pm}$, $B_{+}$, $\mathcal{G}$ and $E_+ + \mathcal{G}$ for $V_{\mathcal{E}}>|V_c|$ (first row) and $V_{\mathcal{E}}<|V_c|$ (second row), respectively. The color coding in $E_{\pm}$ represents the out-of-plane pseudo-spin $\braket{\sigma_z}_\pm$ distribution, where one can clearly identify the occurrence of topological band inversion when $V_{\mathcal{E}}<|V_c|$. Clearly, both the intensity and profile of $B_{+}$ and $\mathcal{G}$ change dramatically with interlayer bias. The profiles in Fig.~\ref{Fig:E_and_B_G_vs_bias}(e) can be understood as originating from Fig.~\ref{Fig:PseudospinAndMagneticField2deg}(d) and Fig.~\ref{Fig:EandGTwist}(b) with the hot spots pushed towards the center of the lower half cell. It is clear that the hot spots are more concentrated and they can be confined to a very localized region with dramatically magnified intensity (e.g. several orders of magnitude) by increasing the bias further. This makes TMD moir\'e applicable for studying the Aharonov-Bohm effect, e.g. by mapping the interference pattern in the local density of states with scanning tunneling microscopy.\cite{ABeffectLDOSNatPhys2011,ABeffectLDOSPRB2009} By increasing the bias beyond the critical value, the magnetic flux becomes zero, indicating that the field exhibits both positive and negative signs in the moir\'e unit cell (Fig.~\ref{Fig:E_and_B_G_vs_bias}(b)). In this regime, the intensity of $B_{+}$ and $\mathcal{G}$ will drop with the increase of bias. Therefore, bias slightly above $|V_c|$ is preferable if fields with high intensities are desirable. Figs.~\ref{Fig:E_and_B_G_vs_bias}(c, f) present the results of $E_+ + \mathcal{G}$, where one can clearly see that the trapping sites for holes (dark red spots) form a triangular lattice (e.g. white triangles in panel (c)). Furthermore, the magnetic flux piercing through adjacent triangles are very different as the field is more localized around locations that overlap with the blue spots. Consequently, interlayer bias can potentially be employed to form effective lattice structures with distinct hopping phases in the tight-binding limit as compared to results in Figs.~\ref{Fig:EandGTwist}(c, d).

\subsection{Strain engineering}\label{Sect:StrainEngineering}
In this section, we discuss the effects of strain and show how strain can be incorporated to manipulate the properties of moir\'e. Here strain will play two roles: (i) It will change the local atomic registries in the moir\'e. (ii) It introduces a pseudo-gauge potential that modifies the phase of interlayer coupling potential $\tilde{U}_{vv}$.

Imagine the top layer is strained after twisting (if there is any). The strain operation is described by the matrix $S=\mathbbm{1}+\epsilon$, where  $\epsilon$ is the strain tensor.
The displacement vector, moir\'e primitive vectors, and reciprocal lattice vectors are then given by $\bdelta(\br)=\left(\mathbbm{1}-R^{-1}S^{-1}\right)\br$, $\bL_i=\left(\mathbbm{1}-R^{-1}S^{-1}\right)^{-1}\ba_i$, and $\bG_i=\left(\mathbbm{1}-S^{-1}R\right)\bb_i$,
respectively.\cite{NanotubeMoirePRB2015} By setting $R=\mathbbm{1}$, one can study the pure effects of strain without rotation. 
Again, one can write $\bG_i=\bb_i-\tilde{\bb}_i$, where $\tilde{\bb}_i=S^{-1}R\bb_i$ is the reciprocal lattice vector of the top manipulated layer. The $K$ points of the top layer now reads $\tilde{\bK}_{\tau}=\tau\left(2\tilde{\bb}_1+\tilde{\bb}_2\right)/3=S^{-1}R\bK_{\tau}$.

In contrast to twisting, strain can also modify the intralayer hopping energy due to variations in the atomic distance in the strained layer.\cite{StrainPhysRep2016,StrainPhysRep2010,ZhaiStrainMPLB} This effect can be described by a pseudo-gauge potential and incorporated as an extra shift of the Dirac points $\tilde{\bK}_{\tau}\rightarrow\tilde{\bK}_{\tau}+\bA^{\tau}_{\text{strain}}/\hbar$, where
\begin{equation}
	\bA^{\tau}_{\text{strain}}=\tau\frac{\sqrt{3}\hbar\beta}{2a}(\epsilon_{xx}-\epsilon_{yy},-2\epsilon_{xy})
\end{equation}
exhibits opposite signs in the two valleys to respect time-reversal symmetry, and $\beta\approx2-3$.\cite{LiangFuStrainMoire,StrainPhysRep2016,StrainPhysRep2010,ZhaiStrainMPLB} This effect has been neglected in the previous studies,\cite{HongyiPseudoFieldMoire} here we will show that it can dramatically modify the in-plane pseudo-spin texture and the resultant geometric magnetic field and scalar potentials. Gap size and band edge energy of the strained layer are also modulated by strain.\cite{StrainedTMDShiangFangPRB2018} We neglect such changes in the following and focus on effects caused by $\bA^{\tau}_{\text{strain}}$. We elaborate on how such extra modulations can be accounted in our approach in Appendix~\ref{App:StrainInducedGapModulation}.

\begin{table*}
	\caption{Geometric properties of various moir\'e patterns formed by small rotation or strain. Note that Poisson's ratio is set to unity in the case of uniaxial strain. Employing a realistic value will compress the lattice along one of the directions as shown in Appendix~\ref{App:StrainInducedGapModulation}. Monolayer primitive vectors are chosen as $\ba_1=\left(1,0\right)a$ and $\ba_2=\left(1/2,\sqrt{3}/2\right)a$. The corresponding monolayer reciprocal lattice vectors are $\bb_1=(1,\,-1/\sqrt{3})2\pi/a$ and $\bb_2=(0,\,2/\sqrt{3})2\pi/a$.}\label{Table}
	
	\begin{ruledtabular}
		\centering
		\begin{tabular}{M{1.3in}|M{1.3in}|M{1.3in}|M{1.3in}|M{1.3in}}
			& Twisting & Biaxial & Uniaxial & Shear \\ [0.05in]
			\hline
			
			$R$ or $S$ & 
			
			$R=\begin{pmatrix}
			\cos \theta&-\sin\theta\\
			\sin\theta&\cos\theta 
			\end{pmatrix}$ &
			
			$S=\begin{pmatrix}
			1+\eta&0\\
			0&1+\eta
			\end{pmatrix}$ & 
			
			$S=\begin{pmatrix}
			1+\eta&0\\
			0&1-\eta
			\end{pmatrix}$ & 
			
			$S=\begin{pmatrix}
			1&\eta\\
			\eta&1
			\end{pmatrix}$ \\ [0.12in]
			\hline
			
			moir\'e reciprocal lattice vector $\bG_i$
			& $\bG_i\approx(\theta b_{i,y},\,-\theta b_{i,x})$ 
			& $\bG_i\approx(\eta b_{i,x},\,\eta b_{i,y})$ 
			& $\bG_i\approx(\eta b_{i,x},\,-\eta b_{i,y})$ 
			& $\bG_i\approx(\eta b_{i,y},\,\eta b_{i,x})$ \\ [0.05in]
			\hline
			
			moir\'e primitive vector
			length $L$
			& $a/\theta$ & $a/\eta$ & $a/\eta$ & $a/\eta$ \\ [0.05in]
			\hline
			
			moir\'e primitive vector $\bL_i$
			&  
			$\begin{aligned}
			\bL_1&\approx(0,-1)L\\
			\bL_2&\approx(\sqrt{3}/2,-1/2)L
			\end{aligned}$
			&  
			$\begin{aligned}
			\bL_1&\approx(1,0)L\\
			\bL_2&\approx(1/2,\sqrt{3}/2)L
			\end{aligned}$
			&
			$\begin{aligned}
			\bL_1&\approx(1,0)L\\
			\bL_2&\approx(1/2,-\sqrt{3}/2)L
			\end{aligned}$ & 
			
			$\begin{aligned}
			\bL_1&\approx(0,1)L\\
			\bL_2&\approx(\sqrt{3}/2,1/2)L
			\end{aligned}$\\ [0.15in]
			\hline
			
			Schematics of $\bL_i$ 
			& \includegraphics[height=0.5in]{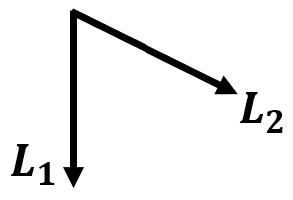} 
			& \includegraphics[height=0.5in]{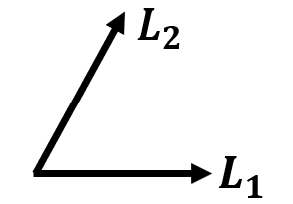}
			& \includegraphics[height=0.5in]{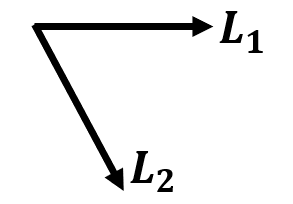}
			& \includegraphics[height=0.5in]{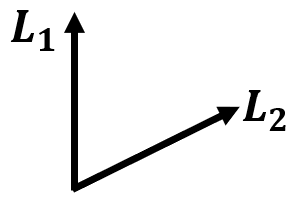} \\
			\hline
			
			$(\epsilon_{xx}-\epsilon_{yy},\,-2\epsilon_{xy})$ 
			& 0 
			& 0
			& $(2\eta,\,0)$
			& $(0,\,-2\eta)$ \\ [0.05in]	
		\end{tabular}
	\end{ruledtabular}
\end{table*}

We will consider simple strain profiles such that components of $\epsilon$ are constant (Table~\ref{Table}). Also. we choose the $\tau=+$ valley and neglect the valley index in the following. As $\bA_{\text{strain}}$ causes an effective shift in $\tilde{\bK}$, the interlayer coupling in Eq.~(\ref{Eq:TildeUvv}) gains a strain-dependent phase, i.e. $\tilde{U}_{vv}\rightarrow\tilde{U}_{vv}e^{-i\bA_{\text{strain}}\cdot\br/\hbar}$. As we have discussed previously, the phase of $\tilde{U}_{vv}$ determines the in-plane pseudo-spin orientation, this explains the mechanism of using strain to engineer the geometric magnetic field and scalar potentials. Also note that $\bA_{\text{strain}}$ is a pure intralayer effect and independent of interlayer registry, information of the latter is contained in the rest of the terms in $\mathcal{U}_v$. 

\begin{figure}[ht]
	\includegraphics[width=3.4in]{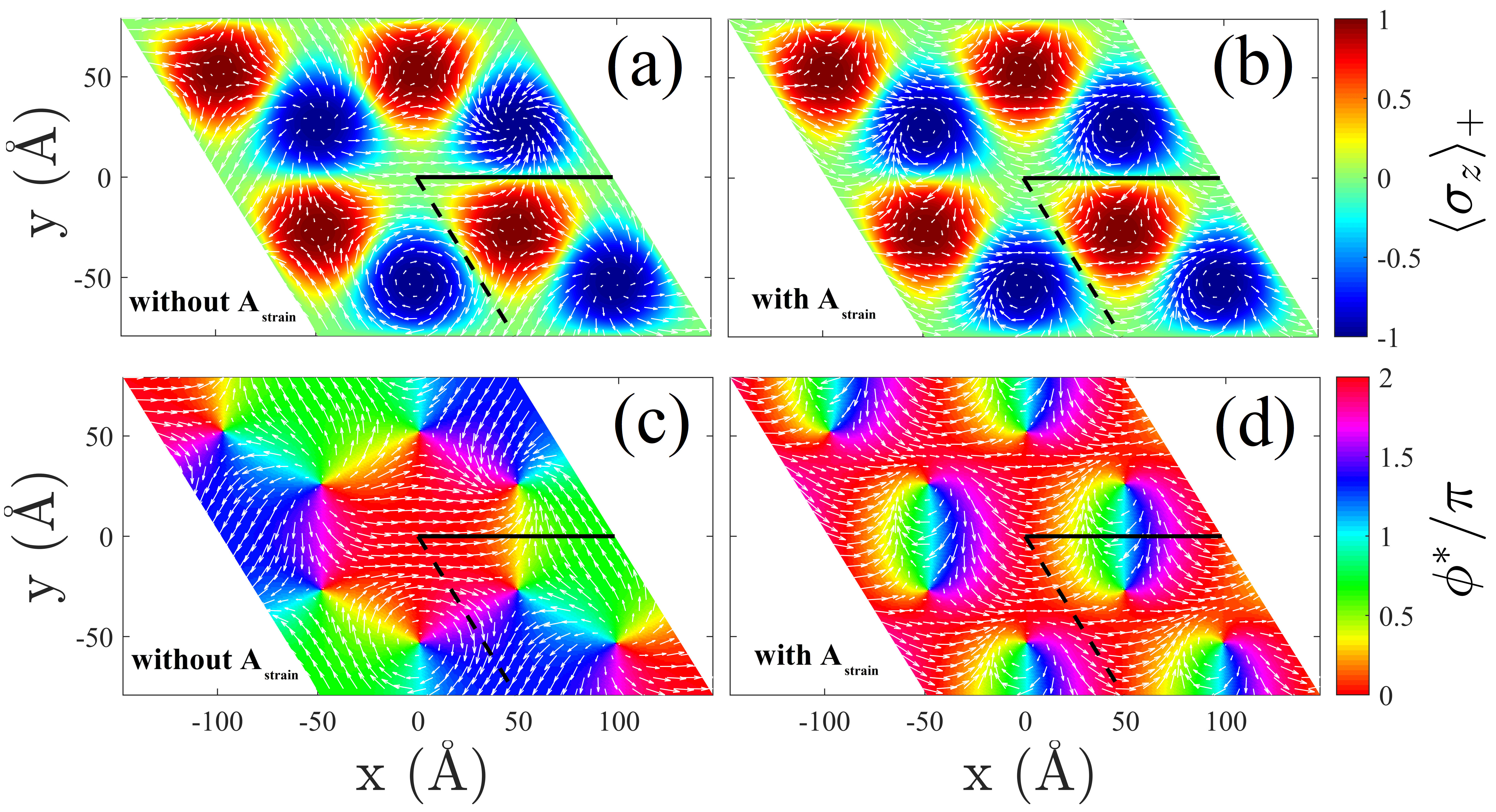}
	\caption{Pseudo-spin distribution for a uniaxially strained bilayer ($\eta=0.035$, $\theta=0$). The first (second) column shows results without (with) the phase correction due to $\bA_{\text{strain}}$. (a, b) The arrows and background color represent in-place and out-of-plane pseudo-spin, respectively. (c, d) The arrows and background color represent in-place pseudo-spin and its orientation, respectively. The solid and dashed lines represent $\bL_1$ and $\bL_2$ (also see schematics in Table~\ref{Table}). $\beta=2.5$ is used here and throughout the rest of the paper unless stated otherwise.}
	\label{Fig:UniaxialPseudoSpin}
\end{figure}

The first (second) column of Fig.~\ref{Fig:UniaxialPseudoSpin} shows the pseudo-spin $\braket{\vec{\sigma}}_{+}$ distribution in the absence (presence) of $\bA_{\text{strain}}$ caused by a uniaxial strain (no twisting is applied). By comparing the two columns, one can clearly see that strain dramatically changes the distribution of in-plane pseudo-spin, leaving $\braket{\sigma_{z}}_+$ unaffected (also same as that from twisting in Fig.~\ref{Fig:PseudospinAndMagneticField2deg}(a)). Most prominently, orientation of $\braket{\bsigma}_+$ exhibits disconnected parallel arrays of circular regions (Fig.~\ref{Fig:UniaxialPseudoSpin}(d) green/blue regions), in contrast to connected hexagonal patches (Fig.~\ref{Fig:UniaxialPseudoSpin}(c)). Furthermore, one may notice that the vorticity of $\braket{\bsigma}_+$ flips sign as compared to the case of twisting in Fig.~\ref{Fig:PseudospinAndMagneticField2deg}(a). This implies an inverse of the magnetic field direction as well as the magnetic flux.

To visualize the effects of strain on the geometric magnetic field and scalar potentials, we will consider three types of strain in the following, i.e. biaxial tensile, zero-average uniaxial, and shear strain. Table~\ref{Table} lists some of their properties in the small strain limit, the case of twisting is also provided for comparison. First we notice that the triangular Bravais lattices (denoted by $\Delta$ in the following) defined by $\bL_{i=1,2}$ in these configurations are related by different symmetries (see the schematics in Table~\ref{Table}): $\Delta_{\text{twist}}$ and $\Delta_{\text{biaxial}}$ are related by $C_{4z}$, $\Delta_{\text{biaxial}}$ and $\Delta_{\text{uniaxial}}$ are related by $C_{2x}$, and $\Delta_{\text{twist}}$ and $\Delta_{\text{shear}}$ are related by $C_{2x}$. A direct consequence of these symmetry relations is, if we neglect the effect of $\bA_{\text{strain}}$ for now and choose $\theta=\eta$,
\begin{equation}
B^{\text{twist}}(\br)=B^{\text{biaxial}}(\br)=-B^{\text{uniaxial}}(\br)=-B^{\text{shear}}(\br)\label{Eq:BFieldRelation}
\end{equation}
in the moir\'e unit cells. The sign reversal in the last two cases is caused by the two-fold rotation around the $x$ axis. 
In practice, one can tell the direction of $\bB_{+}$ simply by looking at the direction defined by $\bL_1\times\bL_2$, with $\bL_i$ defined using the convention $\bL_i=\left(\mathbbm{1}-R^{-1}S^{-1}\right)^{-1}\ba_i$. It should be pointed out that Eq.~(\ref{Eq:BFieldRelation}) is only approximately correct in the small twist and strain limit, where $\bL_i$ and $\bG_i$ exhibit the simple expressions in Table~\ref{Table}. In general, different moir\'e patterns will exhibit distinct orientations and periods, thus magnetic fields in different configurations have slightly distinct intensities. 

\begin{figure*}[ht]
	\includegraphics[width=6in]{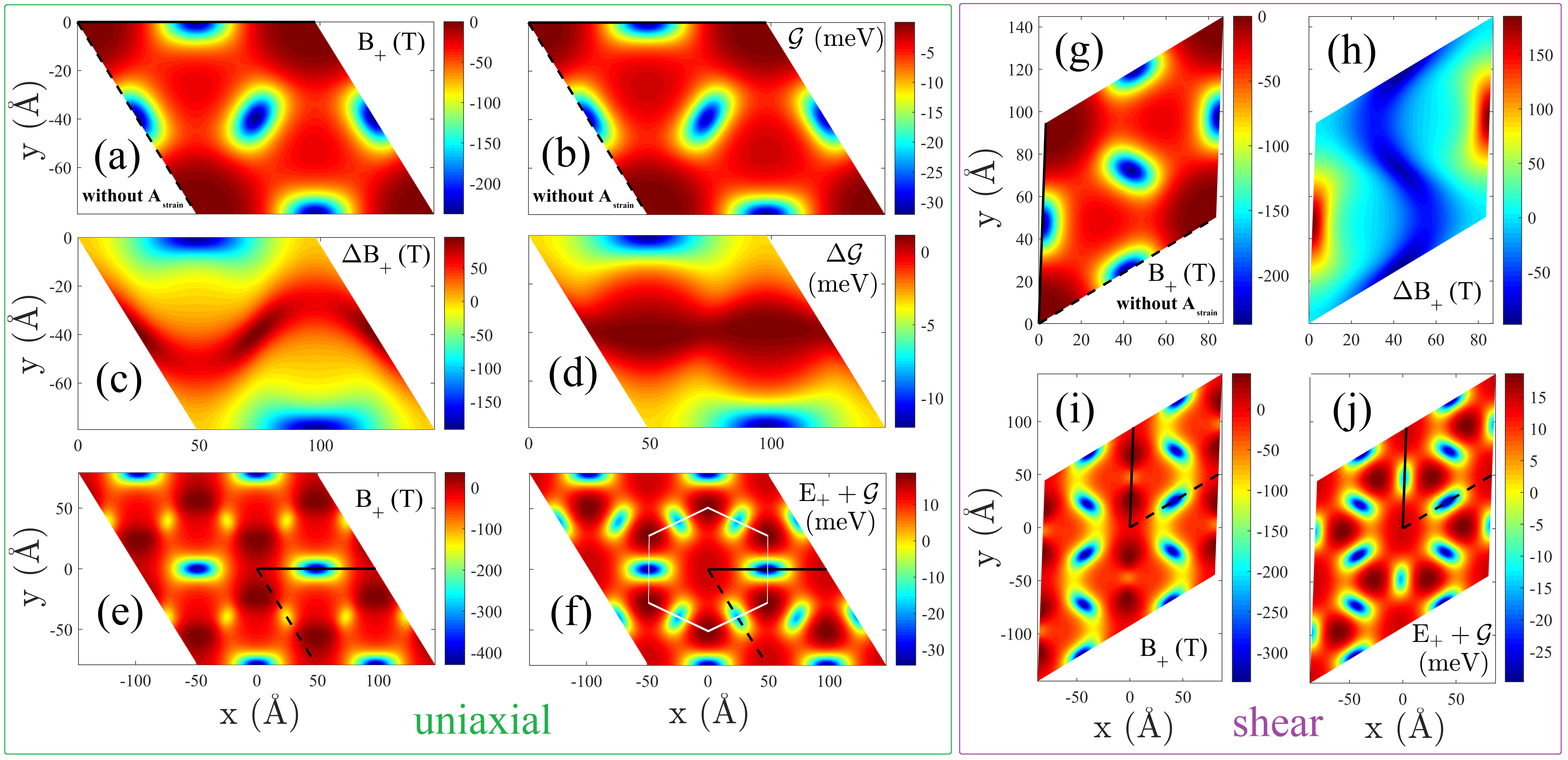}
	\caption{Spatial dependence of magnetic field and scalar potentials in moir\'e structures formed by uniaxial (green box) and shear (purple box) strain ($\eta=0.035$). (a, b) $B_+$ and $\mathcal{G}$ without the effect of $\bA_{\text{strain}}$ in one moir\'e unit cell formed by uniaxial strain. The solid and dashed lines represent $\bL_1$ and $\bL_2$, respectively. (c, d) Net effects $\Delta B_{+}$ and $\Delta \mathcal{G}$ caused by $\bA_{\text{strain}}$. (e, f) $B_+$ and $E_++\mathcal{G}$ with the effect of $\bA_{\text{strain}}$ in four moir\'e unit cells. (g--j) Similar results in the case of shear strain.}
	\label{Fig:Uniaxial_and_Shear_BG}
\end{figure*}

Figs.~\ref{Fig:Uniaxial_and_Shear_BG}(a, b) present the numerical results of $B_{+}$ and $\mathcal{G}$ in the case of uniaxial strain without $\bA_{\text{strain}}$. As discussed in the above, the sign of the magnetic field is reversed (compare Figs.~\ref{Fig:Uniaxial_and_Shear_BG}(a) and \ref{Fig:PseudospinAndMagneticField2deg}(d)). One can check that the profile of Fig.~\ref{Fig:Uniaxial_and_Shear_BG}(a) indeed can be obtained from Fig.~\ref{Fig:PseudospinAndMagneticField2deg}(d) via $C_{4z}$ followed by $C_{2x}$. The results for biaxial strain is a trivial rotation of Fig.~\ref{Fig:PseudospinAndMagneticField2deg}(d) around the $z$ axis (not shown). Fig.~\ref{Fig:Uniaxial_and_Shear_BG}(g) shows the results of $B_+$ in the presence of shear strain, which can be obtained from Fig.~\ref{Fig:PseudospinAndMagneticField2deg}(d) by $C_{2x}$.
The arguments of obtaining the magnetic flux based on the solid angle enclosed by the pseudo-spin on the Bloch sphere remains valid. Therefore, magnetic flux is always quantized at one flux quantum independent of the origin of the moir\'e (twisting, strain, or even combination of both).

Now let us look at how the inclusion of $\bA_{\text{strain}}$ affects $\bB_{+}$ and $\mathcal{G}$. First, we notice that $\bA^{\text{biaxial}}_{\text{strain}}=0$. Since $\Delta_{\text{twist}}$ and $\Delta_{\text{biaxial}}$ are related by a trivial rotation, we conclude that biaxial strain can be employed to replace twisting, which could be experimentally challenging to tune at small angles, to achieve moir\'e structures with similar physical properties. In contrast, uniaxial and shear strain induce non-zero $\bA_{\text{strain}}$, which satisfy $\bA^{\text{uniaxial}}_{\text{strain}}\propto(2\eta,\,0)$ and $\bA^{\text{shear}}_{\text{strain}}\propto(0,\,-2\eta)$, respectively.
Recall that $\bB_{+}$ and $\mathcal{G}$ are given by Eqs.~(\ref{Eq:B+_rigid_twist}) and (\ref{Eq:G_rigid_twisting}), and the azimuthal angle is modified according to $\phi^{*}\rightarrow\phi^{*}+\bA_{\text{strain}}\cdot\br/\hbar$. One can obtain the change of $\bB_{+}$ and $\mathcal{G}$ as
\begin{equation}
	\begin{aligned}
	\Delta \bB_{+}&=-\frac{\Phi_0}{4\pi\hbar}\nabla\left(\cos\zeta\right)\times\bA_{\text{strain}}\\
	\Delta\mathcal{G}&=-\frac{\sin^2\zeta}{8m^*}\left(\bA^2_{\text{strain}}+2\hbar\bA_{\text{strain}}\cdot\nabla\phi^{*}\right)
	\end{aligned},\label{Eq:DeltaB_and_DeltaG_constant_strain}
\end{equation}
where $\phi^{*}$ in the second line represents the phase of $\tilde{U}^{*}_{vv}$ without the contribution of $\bA_{\text{strain}}$.
Figs.~\ref{Fig:Uniaxial_and_Shear_BG}(c, d) show $\Delta B_{+}$ and $\Delta \mathcal{G}$ due to $\bA^{\text{uniaxial}}_{\text{strain}}$. One can clearly identify the intense minima (blue spots) on the top and bottom edges, as well as more extended maxima (red area) with weaker intensity occurring in the middle.
Figs.~\ref{Fig:Uniaxial_and_Shear_BG}(e, f) present the results of $B_{+}$ and $E_++\mathcal{G}$ in the presence of $\bA^{\text{uniaxial}}_{\text{strain}}$ (profile of $\mathcal{G}$ is similar to that of $B_+$, so not shown). Note that four moir\'e unit cells are included to illustrate the breaking of the three-fold rotational symmetry that is present in the case of twisting. The minima occurring on the top and bottom edges of the unit cell boundaries are enhanced. More prominently, features along the central horizontal direction are weakened because $\phi^{*}$ varies more slowly there (also refer to the background color of Fig.~\ref{Fig:UniaxialPseudoSpin}(d)). Figs.~\ref{Fig:Uniaxial_and_Shear_BG}(h--j) present the effects of shear strain (results of $\mathcal{G}$ not shown due to similarity to $B_+$). By comparing Figs.~\ref{Fig:Uniaxial_and_Shear_BG}(h) and (c), one can see that the sign of $\Delta B_+$ is reversed apart from a rotation around the $z$ axis. This makes the fields exhibit zigzag stripes in the $y$ direction (blue spots in Figs.~\ref{Fig:Uniaxial_and_Shear_BG}(i, j)). By inspecting the profiles of the scalar potentials in Figs.~\ref{Fig:Uniaxial_and_Shear_BG}(f, j), one can see that the hopping energy distribution between trapping sites for holes in the tight-binding limit is more complicated in the presence of strain (e.g. schematically represented by white lines with different widths in panel (f)) as the blue barriers in different directions have distinct magnitudes. One may refer to Ref.~\cite{LiangFuStrainMoire} (e.g. Figs. 9 and 10 therein) for electronic properties of moir\'e formed by strain. As both the magnitude and phase of the hopping are modulated by the strain-induced rearrangement of the magnetic field and scalar potential, one might expect topological phase transitions as the strength or direction of strain is manipulated. For instance, Fig. 10(c) of Ref.~\cite{LiangFuStrainMoire} shows that moir\'e formed by uniaxial/shear strain (corresponds to $\varphi=0^\circ/90^\circ$ therein) is topologically trivial/non-trivial. Furthermore, the topological properties of the system can be switched from trivial to non-trivial periodically by continuously tuning the direction of the strain.\cite{LiangFuStrainMoire} The strain-dependent landscapes of magnetic field and scalar potential in our work shed light on the physical origin of these numerical findings.

The above observations suggest that strain can be utilized to tune both the magnitude and profile of the fields via in-plane pseudo-spin engineering. Further tuning can be achieved by varying the size and direction of the strain, or by making combinations of different strain types. For example, $\bA_{\text{strain}}\propto(2\eta,\,-2\eta)$ can be achieved by combining uniaxial and shear strain. From the robustness of skyrmion type structure of the pseudo-spin distribution (e.g. Fig.~\ref{Fig:UniaxialPseudoSpin}(b)), one expects that $\Delta B_{+}$ does not affect the magnetic flux. This can also be confirmed mathematically by noticing that $\Delta \bB_{+}\propto\nabla\left(\cos\zeta\right)\times\bA_{\text{strain}}=\nabla\times\left(\bA_{\text{strain}}\cos\zeta\right)$. With $\bA_{\text{strain}}\cos\zeta$ being well-defined inside the moir\'e unit cell, one can easily see that its integral along the moir\'e boundaries vanishes. So it has no contribution to the magnetic flux according to Stokes' theorem. On the other hand, strain will affect the `flux' of the geometric scalar potential $\mathcal{G}$. We find that the surface integral of $\mathcal{G}$ is only conserved against variation in strain intensity when $|\bL_1|=|\bL_2|$ (e.g. in the case of biaxial strain) even when $\bA_{\text{strain}}$ is excluded.

\begin{figure}[ht]
	\includegraphics[width=3.4in]{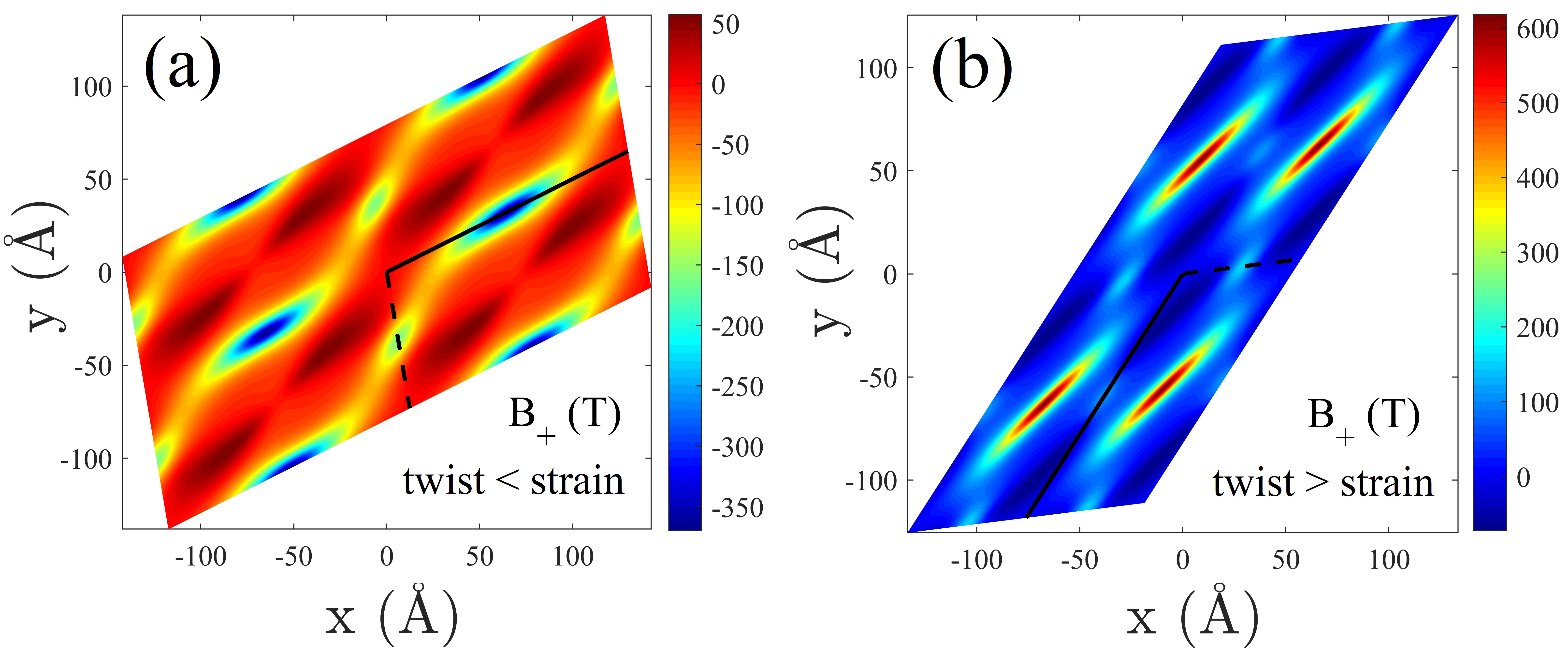}
	\caption{Spatial dependence of $B_{+}$ in the presence of both twisting and uniaxial strain. (a) $\theta=1^\circ$, $\eta=0.035$. (b) $\theta=3^\circ$, $\eta=0.035$. Note that $2^\circ$ is equivalent to $0.035$ in magnitude.}
	\label{Fig:TwistUniaxialCompete}
\end{figure}

As another example, we consider the coexistence of twisting and strain to illustrate how they compete. Fig.~\ref{Fig:TwistUniaxialCompete} presents $B_+$ in the presence of uniaxial strain with fixed strength, while twisting is applied with different angles. In panel (a), strain intensity is stronger than the twisting, thus the profile is a distorted version of Fig.~\ref{Fig:Uniaxial_and_Shear_BG}(e). The maximum intensity is weaker because of the partial cancellation from the effect of twisting, which contributes magnetic field in the opposite direction (see Eq.~(\ref{Eq:BFieldRelation})). In contrast, the effect of twisting is stronger in panel (b), which results in positive field hot spots in the middle of every unit cell. Weaker magnetic field spots (cyan regions) also exist on the boundaries of the unit cells. Overall, the profile resembles that in Fig.~\ref{Fig:PseudospinAndMagneticField2deg}(d), although it becomes more irregular.

Finally, we want to stress that $\bA_{\text{strain}}$ due to simple constant $\epsilon$ (e.g. those in Table~\ref{Table}) does not induce any magnetic field in the case of monolayers because $\nabla\times\bA_{\text{strain}}\equiv0$.\cite{StrainPhysRep2016,StrainPhysRep2010,SasakiStrainGraphene,ZhaiStrainMPLB} This clearly demonstrates the important role of layer pseudo-spin internal DoF and its non-trivial spatial texture in the emergence of the moir\'e magnetic field discussed above. Another difference between the non-uniform strain-induced pseudo-magnetic field in 2D materials (without moir\'e) and the emergent geometric magnetic field in moir\'e is that the flux in the former case is vanishing,\cite{SasakiStrainGraphene} while it could be non-zero and quantized in the latter as discussed in this work. It is also interesting to compare with the situation of twisted bilayer graphene. Due to the gapless nature of graphene, one cannot eliminate the conduction band and only focus on the valence band as for TMD. However, one can still define a gauge potential within the four-band model in twisted bilayer graphene.\cite{NonAbelianGrapheneBilayerPRL2012,PseudoFieldTwistedBilayerDaiXi} Only at the AA stacking locals can the gauge structure in twisted bilayer graphene be simplified to a form equivalent to that of a uniform pseudo-magnetic field $\propto L^{-1}$, where $L$ is the moir\'e period.\cite{PseudoFieldTwistedBilayerDaiXi} In contrast, the pseudo-magnetic field in our case is inhomogeneous with the $L^{-2}$ scaling, and the pseudo-magnetic field description is applicable in the entire moir\'e lattice.

One may also wonder about the possibility of observing Landau levels (LLs) in the moir\'e lattice. For well-defined LLs to exist, the magnetic field should be uniform at the scale of magnetic length. Therefore, it is required that $l\gg l_B$, where $l$ denotes the length scale over which the magnetic field $B$ roughly stays constant, and $l_B=\sqrt{\hbar/eB}$ is the magnetic length.\cite{LLConditionPRB2016,LLConditionSciAdv2019} Equivalently, the condition can be written as $Bl^2/\Phi_0\gg1$. It is clear from Fig.~\ref{Fig:PseudospinAndMagneticField2deg}(d) that the magnetic field is highly non-uniform, and the flux through a unit cell is $\Phi_0$. If we treat a red spot in Fig.~\ref{Fig:PseudospinAndMagneticField2deg}(d) as a region where the field is roughly uniform, one can see that $Bl^2/\Phi_0<0.5$, hence well-defined LLs are not expected to emerge. While LL physics is not relevant here, this pseudo-magnetic field profile realizes fluxed superlattices, for example, the Haldane model for quantum anomalous/spin Hall effect.\cite{WuMacDonaldPRL2019,HongyiPseudoFieldMoire}

Finally, we want to comment on the effects of spontaneous lattice relaxation in the moir\'e. Due to the presence of various local stacking configurations, among which some are energetically unfavorable, lattice relaxation becomes prominent when the twist angle is small. Relaxation will expand the area of $R^X_M$ and $R^M_X$ stacking locals, while $R^A_A$ stacking regions will shrink, and narrow solitons will form at domain boundaries.\cite{FalkoRelaxationTheory,FalkoRelaxationExpt,RelaxationTMDJPCC,FlatBandSolitonTwistedTMDPRL2018} 
As to the pseudo-magnetic field $\bB_{+}$, its profile can be changed by the spontaneous relaxation strain, but the magnetic flux per unit cell corresponds to the solid angle covered by layer pseudo-spin, whose quantized value is unmodified.
On the other hand, compared to rigidly twisted bilayers, the moir\'e-modulated interlayer coupling changes more rapidly near the solitons, where stronger coupling between the two energy branches $E_\pm$ and a larger geometric scalar potential correction can be expected. These can affect the validity of the adiabatic approximation near the solitons. However, we expect that spontaneous lattice relaxation is quenched in moir\'e formed by external strain, and one can also reduce the effects of lattice relaxation in twisted bilayers with clamped edges.

\section{Summary}
To summarize, we have shown that moir\'e patterns spatially modulate the layer pseudo-spin in a homobilayer TMD. When particles undergo an adiabatic evolution, their dynamics are governed by a geometric magnetic field and a scalar potential. The profile and intensity of the magnetic field are controllable by varying the twist angle, interlayer bias, and strain. The magnetic flux per moir\'e unit cell is quantized and tunable with interlayer bias.
The landscape of the scalar potential is also sensitive to the above tuning knobs, which forms various effective tight-binding lattice structures. We expect that such tunable flux lattices built from moir\'e patterns are promising for exploring valley/spintronics and topological properties.

\section{Outlook}
Geometric fields and potentials originated from spatial textures have profound effects on various aspects of materials.\cite{QianNiuRMP,GoldmanNonAbelianReview,NonAbelianGaugeRMP} This work studies the real space manifestation of such effects in homobilayer TMD moir\'e lattices as a pseudo-magnetic field and a geometric scalar potential. A systematic exploration of the tunability of their landscape via twist angle, interlayer bias, and uniform strain is presented. Our results serve as guides for experimental studies as well as shed light on the physical origin of variations in the electronic and topological properties of moir\'e patterns under different conditions.\cite{LiangFuStrainMoire,WuMacDonaldPRL2019} For future studies, one can incorporate spatially nonuniform strain (e.g. via substrate engineering or intrinsic strain caused by spontaneous lattice relaxation) into moir\'e patterns, which may add further spatial tunability to the existing results. Possible valley/spin polarized phenomena in transport can also be investigated by combining moir\'e pseudo-magnetic field (with valley contrasted sign) and external magnetic field (valley independent). It is also interesting to explore situations where non-adiabatic effects emerge. To do so, one can, for example, explore the higher energy regime (compared to moir\'e interlayer coupling intensity), or add proper interlayer bias to bring the two energy branches in close proximity (Sect.~\ref{Sect:BiasEngineering}). In such scenarios, one should focus on the non-Abelian Berry connection, which is flat without curvature as discussed in the last paragraph of Sect.~\ref{Sect:NonAbelianGauge}.

\begin{acknowledgements}
The authors thank Hongyi Yu for helpful discussions. The work is supported by the Research Grants Council of Hong Kong (Grants No. HKU17306819 and No. C7036-17W), and the University of Hong Kong (Seed Funding for Strategic Interdisciplinary Research).
\end{acknowledgements}

\newpage
\appendix

\section{Four-band continuum model of twisted bilayer TMD}\label{App:FourBandContinuumModel}
For pedagogical purposes, here we present details of the four-band model that takes into account both conduction and valence bands for twisted bilayer TMD.
Within the two-band model that covers the lowest conduction band and highest valence band, the two monolayers can be described by\cite{WuMacDonaldPRL2019,InterlayerCouplingNewJPhys2015,WannierTBG,NanotubeMoirePRB2015,LiangFuStrainMoire}
\begin{equation}
	\begin{aligned}
	H_0^t&=\hbar v_F\left[R^{-1}\left(\bk-\tilde{\bK}_{\tau}\right)\right]\cdot\tilde{\bsigma}+\frac{E_g}{2}\tilde{\sigma}_z\\
	H_0^b&=\hbar v_F \left(\bk-\bK_{\tau}\right)\cdot\tilde{\bsigma}+\frac{E_g}{2}\tilde{\sigma}_z
	\end{aligned},
\end{equation}
where $t$ and $b$ label the top and bottom layer respectively, $E_g$ characterizes the band gap, and $\tilde{\bsigma}=(\tilde{\sigma}_x,\tilde{\sigma}_y)$ and $\tilde{\sigma}_z$ are Pauli matrices defined in the metal $d$ orbitals basis $\left(d_{z^2},\,d_{x^2-y^2}+\tau id_{xy}\right)$. One should not confuse $\tilde{\sigma}$ with $\sigma$ in the main text, the latter is used to denote layer pseudo-spin. Also, here $\bk$ should be understood as the operator $-i\nabla$ that measures the wave vector from the $\Gamma$ point.

In the presence of a large moir\'e, the two valleys are decoupled due to vanishing inter-valley scattering and related by time-reversal symmetry. We will focus on the $\tau=+$ valley in the following and neglect the valley index when no confusion arises.
The Hamiltonian for the coupled bilayer is then modeled by
\begin{widetext}
	\begin{equation}
	\begin{aligned}
	\mathcal{H}&=
	\begin{pmatrix}
	H_0^t+V^t&U\\
	U^{\dagger}&H_0^b+V^b
	\end{pmatrix}\\
	&=
	\begin{pmatrix}
	\hbar v_F\left[R^{-1}\left(\bk-\tilde{\bK}\right)\right]\cdot\tilde{\bsigma}+\frac{E_g}{2}\tilde{\sigma}_z+V^t&U\\
	U^{\dagger}&\hbar v_F\left(\bk-\bK\right)\cdot\tilde{\bsigma}+\frac{E_g}{2}\tilde{\sigma}_z+V^b
	\end{pmatrix}
	\end{aligned},
	\end{equation}
\end{widetext}
where $V^{t,b}$ are diagonal matrices characterizing the band edge changes in each layer due to perturbation from higher energy bands, and $U$ describes interlayer coupling.\cite{WuMacDonaldPRL2019,InterlayerCouplingTMDPRB2017,LiangFuStrainMoire,HongyiPseudoFieldMoire} The interlayer coupling can be estimated from the two-center approximation keeping the leading contributions by taking advantage of the fact that the hopping energy decays fast with momentum. Its explicit form, and obviously the unperturbed Hamiltonian as well, depends on the choice out of the three equivalent $\bK$ points in the Brillouin zone.\cite{InterlayerCouplingNewJPhys2015,InterlayerCouplingTMDPRB2017} However, such a dependence can be eliminated by performing a unitary transformation as will be shown in the following.

One may notice that $\left[R^{-1}\left(\bk-\tilde{\bK}\right)\right]\cdot\tilde{\bsigma}=e^{i\frac{\theta}{2}\tilde{\sigma}_z}\left[\left(\bk-\tilde{\bK}\right)\cdot\tilde{\bsigma}\right]e^{-i\frac{\theta}{2}\tilde{\sigma}_z}$. This suggests that we can apply a unitary transformation $T_{\theta}=\text{diag}\left(e^{-i\frac{\theta}{2}\tilde{\sigma}_z},\,\mathbbm{1}\right)$ to remove the rotation matrix in the top block:
\begin{widetext}
	\begin{equation}
	\begin{aligned}
	\mathcal{H}\rightarrow H^{'}
	&=T_{\theta}\mathcal{H}T_{\theta}^{\dagger}\\
	&=\begin{pmatrix}
	\hbar v_F\left(\bk-\tilde{\bK}\right)\cdot\tilde{\bsigma}+\frac{E_g}{2}\tilde{\sigma}_z+V^t&e^{-i\frac{\theta}{2}\tilde{\sigma}_z}U\\
	U^{\dagger}e^{i\frac{\theta}{2}\tilde{\sigma}_z}&\hbar v_F\left(\bk-\bK\right)\cdot\tilde{\bsigma}+\frac{E_g}{2}\tilde{\sigma}_z+V^b
	\end{pmatrix}
	\end{aligned},
	\end{equation}
\end{widetext}
at the expense of modifying the interlayer coupling ($e^{-i\frac{\theta}{2}\tilde{\sigma}_z}$ adds a phase factor of $e^{\mp i\frac{\theta}{2}}$ to the first and second row of $U$, respectively).
For a specific moir\'e pattern, $\bK$ and $\tilde{\bK}$ are constant vectors, thus another unitary transformation $T_{\bK}=\text{diag}\left(e^{-i\tilde{\bK}\cdot\br},\,e^{-i\bK\cdot\br}\right)$  (here each term is understood as multiplied by the identity matrix $\mathbbm{1}$) can be employed to remove them in the diagonal blocks, i.e.
\begin{widetext}
	\begin{equation}
	\begin{aligned}
	H^{'}\rightarrow H&=T_{\bK}H^{'}T_{\bK}^{\dagger}\\
	&=
	\begin{pmatrix}
	v_F \tilde{\bsigma}\cdot\bp+\frac{E_g}{2}\tilde{\sigma}_z+V^t&\tilde{U}\\
	\tilde{U}^{\dagger}&v_F\tilde{\bsigma}\cdot\bp+\frac{E_g}{2}\tilde{\sigma}_z+V^b
	\end{pmatrix}
	\end{aligned}\label{Eq:Transformation_on_U},
	\end{equation}
\end{widetext}
where we have defined the transformed interlayer coupling $\tilde{U}=e^{-i\frac{\theta}{2}\tilde{\sigma}_z}Ue^{i\left(\bK-\tilde{\bK}\right)\cdot\br}$ and $\bp=\hbar \bk$. The phase $e^{i\left(\bK-\tilde{\bK}\right)\cdot\br}$ lifts the $\bK$ dependence in $U$, making it depend on the nearest-neighbor vectors of the moir\'e reciprocal lattice, where the three-fold rotation symmetry becomes clear. The other phase $e^{-i\frac{\theta}{2}\tilde{\sigma}_z}$, which has no spatial dependence, does not affect the geometric magnetic field and scalar potential discussed in this work.

In the presence of strain, the $K$ points of the top layer and the corresponding monolayer Hamiltonian read \cite{NanotubeMoirePRB2015}
\begin{equation}
\begin{aligned}
\tilde{\bK}_{\tau}&=\tau\left(2\tilde{\bb}_1+\tilde{\bb}_2\right)/3=S^{-1}R\bK_{\tau}\\
H_0^t&=\hbar v_F\left[R^{-1}S\left(\bk-\tilde{\bK}_{\tau}\right)\right]\cdot\tilde{\bsigma}+\frac{E_g}{2}\tilde{\sigma}_z
\end{aligned}.\label{Eq:K_and_H0_in_the_presence_of_strain}
\end{equation}
In experiments, even the ultrahigh strain achievable is merely a few percent, thus one can employ the approximation $S\approx\mathbbm{1}$ and Eq.~(\ref{Eq:Transformation_on_U}) is still valid with the replacement of $\tilde{\bK}$ given in Eq.~(\ref{Eq:K_and_H0_in_the_presence_of_strain}).

\section{Two-band continuum Hamiltonian for the valence bands}\label{App:DeriveTwoBandModel}
Denote the eigenvector of $H$ as $\Psi=\left(\Psi_{tc},\Psi_{tv},\Psi_{bc},\Psi_{bv}\right)^T$, the Schr\"odinger equation can be written in the form of four coupled equations
\begin{widetext}
	\begin{equation}
	\begin{aligned}
	\left(V^{t}_c+\frac{E_g}{2}\right)\Psi_{tc}+v_F\left(p_x-ip_y\right)\Psi_{tv}+\tilde{U}_{cc}\Psi_{bc}+\tilde{U}_{cv}\Psi_{bv}&=E\Psi_{tc}\\
	v_F\left(p_x+ip_y\right)\Psi_{tc}+\left(V^{t}_v-\frac{E_g}{2}\right)\Psi_{tv}+\tilde{U}_{vc}\Psi_{bc}+\tilde{U}_{vv}\Psi_{bv}&=E\Psi_{tv}\\
	\tilde{U}^{*}_{cc}\Psi_{tc}+\tilde{U}^{*}_{vc}\Psi_{tv}+\left(V^{b}_c+\frac{E_g}{2}\right)\Psi_{bc}+v_F\left(p_x-ip_y\right)\Psi_{bv}&=E\Psi_{bc}\\
	\tilde{U}^{*}_{cv}\Psi_{tc}+\tilde{U}^{*}_{vv}\Psi_{tv}+v_F\left(p_x+ip_y\right)\Psi_{bc}+\left(V^{b}_v-\frac{E_g}{2}\right)\Psi_{bv}&=E\Psi_{bv}
	\end{aligned}.
	\end{equation}
\end{widetext}
Since the energy gap is large, it is a good approximation to decouple the conduction and valence bands. In the following, we will focus on the states near the valence band edge, i.e. $E\approx-E_g/2$. The goal is to derive an effective equation describing the valence band states. Compared to the energy gap $E_g$, the various interlayer coupling energies and $E+E_g/2$ are small quantities, thus can be eliminated. To rewrite the conduction band contributions in terms of their valance band counterparts, we will employ the first and third equations, which yield
\begin{equation}
\begin{aligned}
\Psi_{tc}&\approx-\frac{v_F}{E_g}\left(p_x-ip_y\right)\Psi_{tv}\\
\Psi_{bc}&\approx-\frac{v_F}{E_g}\left(p_x-ip_y\right)\Psi_{bv}
\end{aligned},
\end{equation}
respectively. With these two approximations, the second and fourth equations become 
\begin{equation}
\begin{aligned}
-\frac{p^2}{2m^{*}}\Psi_{tv}+\left(V^{t}_v-\frac{E_g}{2}\right)\Psi_{tv}+\tilde{U}_{vv}\Psi_{bv}&=E\Psi_{tv}\\
-\frac{p^2}{2m^{*}}\Psi_{bv}+\left(V^{b}_v-\frac{E_g}{2}\right)\Psi_{bv}+\tilde{U}^{*}_{vv}\Psi_{tv}&=E\Psi_{bv}
\end{aligned},
\end{equation}
where $m^{*}=\frac{E_g}{2v_F^2}$ is the effective mass, and terms associated with $\tilde{U}_{cv}$ and $\tilde{U}_{vc}$ have been discarded due to weak inter-band coupling in the presence of a large gap. One can then easily identify the effective Hamiltonian for the valence bands $H_v$ from the LHS.

\section{Effects of realistic Poisson's ratio and strain induced gap modulation}\label{App:StrainInducedGapModulation}

\begin{figure*}[ht]
	\includegraphics[width=6in]{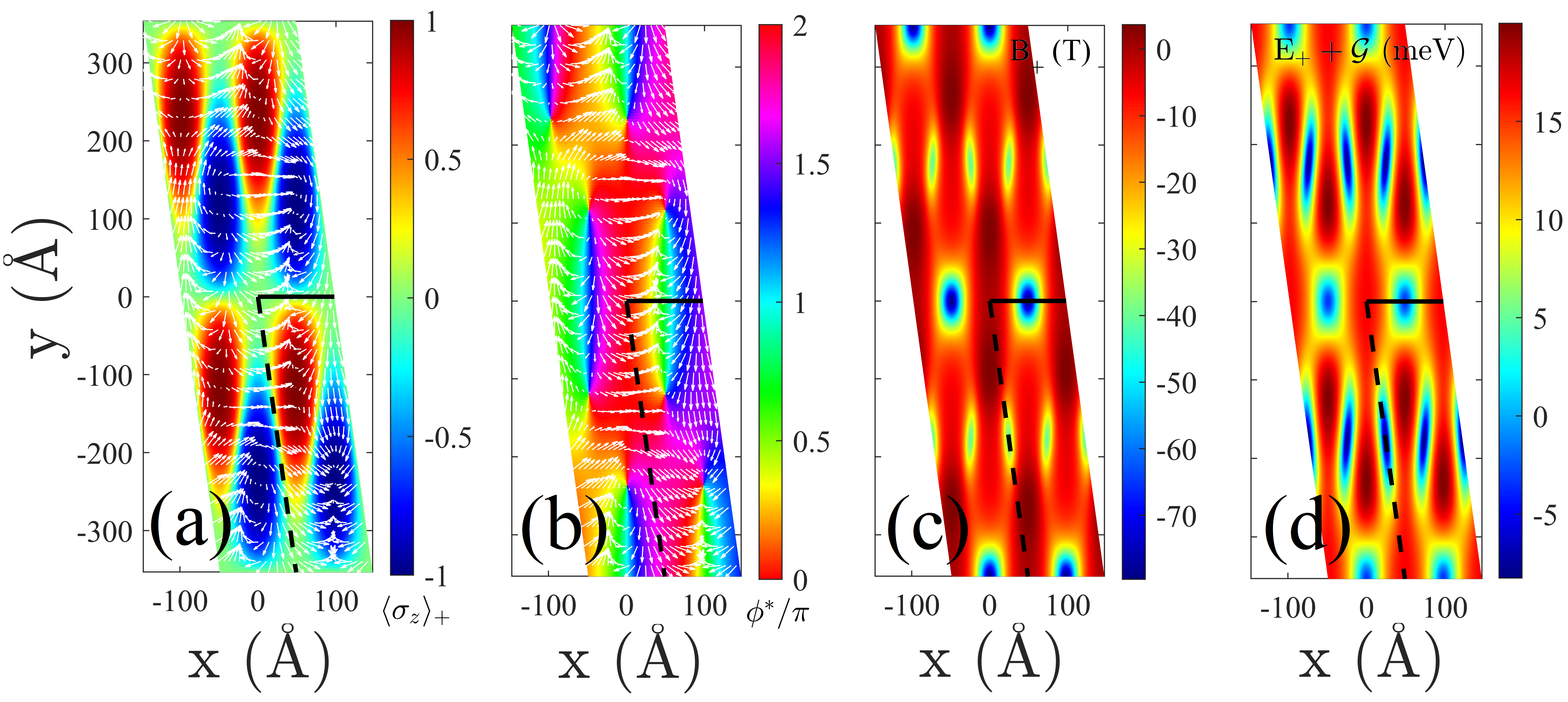}
	\caption{Results for uniaxially strained $(\eta=0.035)$ bilayer MoSe$_2$ with Poisson’s ratio $\nu=0.23$. (a) Layer pseudo-spin distribution, (b) in-plane layer pseudo-spin component and its phase angle $\phi^*$, (c) moir\'e magnetic field $B_+$, and (d) scalar potential $E_++\mathcal{G}$.}
	\label{Fig:UniaxialStrainWithPoissonRatio}
\end{figure*}

In the main text, we have set the Poisson’s ratio $\nu=1$ and neglected the band gap modulation by strain, which allows us to focus on the effects caused by strain-induced pseudo-vector potential. The use of more realistic Poisson's ratio and accounting the band gap modulation by strain can lead to quantitative changes which we explain below.

Employing a Poisson’s ratio $\nu\approx0.23$ for MoSe$_2$,\cite{PoissonRatio} typical results for the case of uniaxial strain are given by those shown in Fig.~\ref{Fig:UniaxialStrainWithPoissonRatio}. The main change is the lattice geometry, i.e. the lattice becomes compressed along one of the directions. The profile of the magnetic field and scalar potential also have quantitative changes. These are because strain tensor components are now given by  $\epsilon=\text{diag}(\eta,\,-\nu\eta)$, different compared to the case of $\nu=1$, leading to distinct moir\'e sizes, interlayer lattice registries, and strain-induced pseudo-vector potentials.

\begin{figure}[ht]
	\includegraphics[width=3.4in]{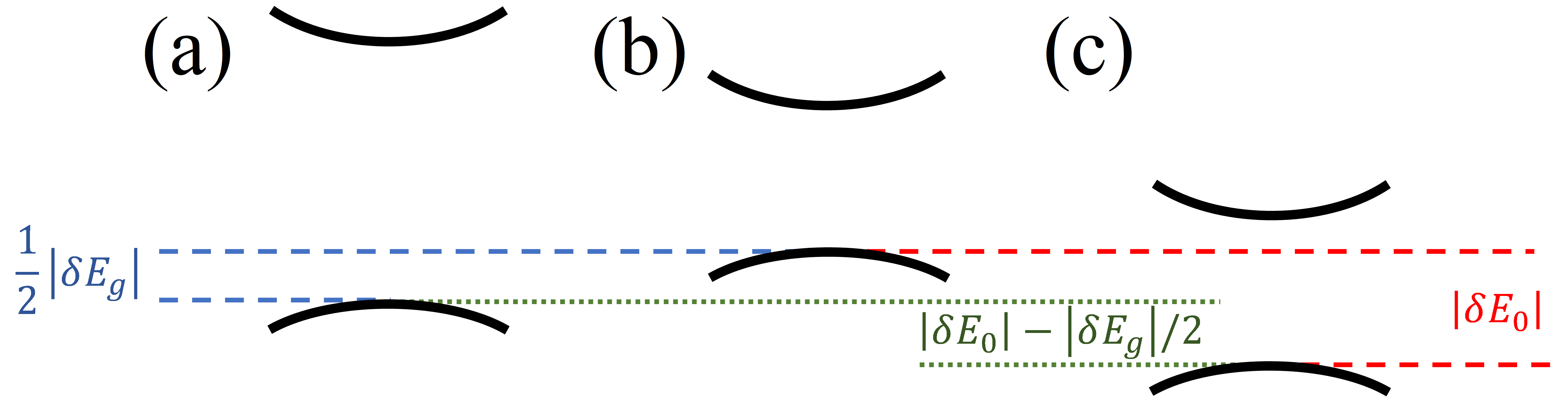}
	\caption{Schematics showing the band gap size and mid-gap position modulated by strain. (a) Bands of unstrained MoSe$_2$. (b) Decrease of band gap by $|\delta E_g|$, hence upward shift of the valence band edge by $|\delta E_g |/2$ (blue dashed lines). (c) Downward shift of the mid-gap position (thus the valence band edge) by $|\delta E_0|$ (red dashed lines). The net offset of the valence band edge caused by strain is $|\delta E_0|-|\delta E_g|/2$ (green dotted lines). Note that energy shifts are exaggerated for clarity.}
	\label{Fig:StrainInducedBandOffset}
\end{figure}

Strain also causes gap size and mid-gap position modulations.\cite{StrainedTMDShiangFangPRB2018} Below we elaborate on how such effect can be accounted in our approach. The gap size modification in the strained layer can be described by $\delta E_g=2\Delta_1(\epsilon_{xx}+\epsilon_{yy})$, where $\Delta_1\approx-2.28$ eV for MoSe$_2$.\cite{StrainedTMDShiangFangPRB2018} In the case of uniaxial strain, it yields $\delta E_g\approx-3.5\eta$ (eV). Additionally, the mid-gap position is also shifted by $\delta E_0=\Delta_2 (\epsilon_{xx}+\epsilon_{yy})\approx-3.85\eta$ (eV), where $\Delta_2\approx-5$ eV for MoSe$_2$.\cite{StrainedTMDShiangFangPRB2018} With both factors considered, the net shift of the valence band edge reads $|\delta E_0|-|\delta E_g|/2\approx2.1\eta$ (eV) (see Fig.~\ref{Fig:StrainInducedBandOffset}). For strain up to a few percent that can be practically achieved, this shift can be up to a few tens of meV, leading to the valence band edge offset between the two layers. In the manuscript, we have discussed the effects of such valence band edge offset introduced by an interlayer bias (Sect.~\ref{Sect:BiasEngineering}). The effects found, i.e. quantitative change in the magnetic field profile, and quantized jump of the magnetic flux at critical value of the band offset, are also applicable when the offset is introduced by strain. On the other hand, one can apply a modest interlayer bias to compensate the band offset caused by strain. The band offset is the control parameter in our discussions that determines the magnetic field profile and flux. This parameter can be contributed by both the strain and interlayer bias, and tunable through the bias at given strain.

In the case of biaxial strain with $\epsilon=\text{diag}(\eta,\,\eta)$, one has $\delta E_g\approx-9\eta$ (eV) and $\delta E_0\approx-10\eta$ (eV), hence the valence band offset reads $|\delta E_0|-|\delta E_g|/2\approx5.5\eta$ (eV). In this case, as strain induced pseudo-vector potential vanishes, one expects the same phenomena as those discussed in Sect.~\ref{Sect:BiasEngineering}.

\newpage
\bibliography{Refs}
\bibliographystyle{apsrev4-1}

\end{document}